\documentclass[prd,amsmath,amssymb,showpacs,superscriptaddress,nofootinbib]{revtex4}
\usepackage{graphicx,epsfig,psfig}
\usepackage{dcolumn}
\usepackage{bm}
\usepackage{rotating}
\usepackage{color}
\usepackage{subfigure}
\usepackage{pstricks}
\usepackage{soul}

\usepackage{overpic}

\def \jpsi      {J/\psi}
\def \psip      {\psi^\prime}

\def \chicj     {\chi_{cJ}}
\def \chiczero  {\chi_{c0}}
\def \chicone   {\chi_{c1}}
\def \chictwo   {\chi_{c2}}
\def \pnpiccp   {p\bar{n}\pi^{-}(\bar{p}n\pi^+)}

\def \pnpim     {p\bar{n}\pi^{-}}
\def \pnpip     {\bar{p}n\pi^{+}}

\def \pnpipim   {p\bar{n}\pi^{-}\pi^{0}}
\def \pnpipip   {\bar{p}n\pi^{+}\pi^{0}}
\def \pppizero  {p\bar{p}\pi^{0}}
\def \pppizeropizero {p\bar{p}\pi^{0}\pi^{0}}
\def \pipizero   {\pi^0\pi^0}
\def \ee         {e^+e^-}
\def \twogamma   {\gamma\gamma}
\def \threegamma {\gamma\gamma\gamma}
\def \fourgamma  {\gamma\gamma\gamma\gamma}

\begin{document}
\normalsize
\parskip=5pt plus 1pt minus 1pt

\title{ \quad\\[1.0cm] \boldmath Measurement of $\chicj$ decaying into
  $\pnpim$ and $\pnpipim$} \author{ M.~Ablikim$^{1}$,
  M.~N.~Achasov$^{5}$, O.~Albayrak$^{3}$, D.~J.~Ambrose$^{39}$,
  F.~F.~An$^{1}$, Q.~An$^{40}$, J.~Z.~Bai$^{1}$, Y.~Ban$^{27}$,
  J.~Becker$^{2}$, J.~V.~Bennett$^{17}$
% ...
  \\
  \vspace{0.2cm}
  (BESIII Collaboration)\\
  \vspace{0.2cm} {\it
    $^{1}$ Institute of High Energy Physics, Beijing 100049, P. R. China\\
    $^{2}$ Bochum Ruhr-University, 44780 Bochum, Germany\\
    $^{3}$ Carnegie Mellon University, Pittsburgh, PA 15213, USA\\
% ...
$^{5}$ G.I. Budker Institute of Nuclear Physics SB RAS (BINP), Novosibirsk 630090, Russia\\
% ...
$^{17}$ Indiana University, Bloomington, Indiana 47405, USA\\
% ...
$^{27}$ Peking University, Beijing 100871, P. R. China\\
% ...
$^{39}$ University of Rochester, Rochester, New York 14627, USA\\
    $^{40}$ University of Science and Technology of China, Hefei 230026, P. R. China\\
% ...
  } } \date{\today}

\begin{abstract}
  Using a data sample of $1.06 \times 10^{8}$ $\psip$ events collected
  with the BESIII detector in 2009, the branching fractions of
  $\chicj\to\pnpim$ and $\chicj\to\pnpipim$ ($J$=0,1,2) are
  measured\footnote{Throughout the text, inclusion of charge conjugate
    modes is implied if not stated otherwise.}. The results for
  $\chiczero\to\pnpim$ and $\chictwo\to\pnpim$ are consistent with,
  but much more precise than those of previous measurements. The
  decays of $\chicone\to\pnpim$ and $\chicj\to\pnpipim$ are observed
  for the first time.
\end{abstract}

\pacs{14.20.Gk, 13.75.Gx, 13.25.Gv}

\maketitle

\section{Introduction}

Exclusive heavy quarkonium decays constitute an important laboratory
for investigating perturbative Quantum Chromodynamics (pQCD). Compared
to $\jpsi$ and $\psip$, relatively little is known concerning $\chicj$
decays~\cite{pdg}. More experimental data on exclusive decays of
P-wave charmonia is important for a better understanding of the nature
of $\chicj$ states, as well as testing QCD based
calculations. Although these states are not directly produced in $\ee$
collisions, they are produced copiously in radiative decays of
$\psip$, with branching fractions around 9\%~\cite{pdg}. The large
$\psip$ data sample taken with the Beijing Spectrometer (BESIII)
located at the Beijing Electron-Positron Collider (BEPCII) provides an
opportunity for a detailed study of $\chicj$ decays.

Previous studies indicate that the Color Octet Mechanism
(COM)~\cite{me}, $c\bar{c}g\to2(q\bar{q})$ could have large effects on
the observed decay patterns of these P-wave charmonia
states~\cite{com1,com2,com3,com4,com5}.  To arrive at a comprehensive
understanding about P-wave dynamics, both theoretical predictions
employing the COM and new precise experimental measurements for
$\chicj$ many-body final states decays are required.

Also, the decays of $\chicj$ with baryons in the final states provide
an excellent place to investigate the production and decay rates of
excited nucleon $N^\ast$ states, which are a very important source of
information for understanding the internal structure of the
nucleon~\cite{nstar1}.

The decays $\chicj\to\pnpim$ and $\chicj\to\pnpipim$ were studied by
the BESII experiment with 14$\times 10^6$ $\psip$
events~\cite{bes2pnpi}. Due to limited statistics and the detector
performance, only $\chi_{c0}\to\pnpim$ and $\chi_{c2}\to\pnpim$ were
observed and measured with large uncertainties. A series of three-body
and four-body decays of $\chicj$, including channels with similar
final states to ours such as $\pppizero$ and $\pppizeropizero$, were
measured by the CLEO-c Collaboration ~\cite{cleo3body,cleo4body}.

In this paper, we present a measurement of $\chicj$ decaying into
$\pnpim$ and $\pnpipim$. The samples used in this analysis consist of
156.4 pb$^{-1}$ of $\psip$ data corresponding to
(1.06$\pm$0.04$)\times 10^8$ ~\cite{totaln} events taken at $\sqrt{s}$
= 3.686 GeV/$c^{2}$, and 42.6 pb$^{-1}$ of continuum data taken at
$\sqrt{s}$ = 3.65 GeV/$c^{2}$.


\section{Detector and Monte-Carlo simulation}

BESIII~\cite{liu2} is a major upgrade of the BESII experiment at the
BEPCII accelerator~\cite{liu3} for studies of hadron spectroscopy as
well as $\tau$-charm physics~\cite{liu4}. The design peak luminosity
of the double-ring $e^{+}e^{-}$ collider, BEPCII, is 10$^{33}$
cm$^{-2}s^{-1}$ at center-of-mass energy of 3.78 GeV. The BESIII
detector with a geometrical acceptance of 93\% of 4$\pi$, consists of
the following main components: 1) a small cell, helium-based main
draft chamber (MDC) with 43 layers. The average single wire resolution
is 135 $\mu$m, and the momentum resolution for 1 GeV/$c$ charged
particles in a 1 T magnetic field is 0.5\%; 2) an electromagnetic
calorimeter (EMC) comprised of 6240 CsI (Tl) crystals arranged in a
cylindrical shape (barrel) plus two end-caps. For 1.0 GeV/$c$ photons,
the energy resolution is 2.5\% (5\%) and the position resolution is 6
mm (9 mm) in the barrel (end-caps); 3) a Time-Of-Flight system (TOF)
for particle identification (PID) composed of a barrel part
constructed of two layers with 88 pieces of 5 cm thick, 2.4 m long
plastic scintillators in each layer, and two end-caps with 48
fan-shaped, 5 cm thick, plastic scintillators in each end-cap. The
time resolution is 80 ps (110 ps) in the barrel (end-caps),
corresponding to a $K/\pi$ separation by more than $2\sigma$ for
momenta below about 1 GeV/$c$; 4) a muon chamber system (MUC)
consisting of 1000 m$^{2}$ of Resistive Plate Chambers (RPC) arranged
in 9 layers in the barrel and 8 layers in the end-caps and
incorporated in the return iron yoke of the superconducting
magnet. The position resolution is about 2 cm.

The optimization of the event selection and the estimation of
backgrounds are performed through Monte Carlo (MC) simulation. The
\textsc{geant}{\footnotesize 4}-based simulation software
\textsc{boost}~\cite{liu5} includes the geometric and material
description of the BESIII detectors and the detector response and
digitization models, as well as the tracking of the detector running
conditions and performance. The production of the $\psip$ resonance is
simulated by the MC event generator \textsc{kkmc}~\cite{liu6}, while
the decays are generated by \textsc{evtgen}~\cite{liu7} for known
decay modes with branching fractions being set to world average
values~\cite{pdg}, and by \textsc{lundcharm}~\cite{liu9} for the
remaining unknown decays.

\section{Event selection}

Charged-particle tracks in the polar angle range $|\cos\theta|<$0.93
are reconstructed from hits in the MDC. Only the tracks with the point
of closest approach to the beamline within $\pm$5 cm of the
interaction point in the beam direction, and within 0.5 cm in the
plane perpendicular to the beam are selected. The TOF and $dE/dx$
information are used to form particle identification (PID) confidence
levels for $\pi$, $K$ and $p$ hypotheses. Each track is assigned to
the particle type that corresponds to the hypothesis with the highest
confidence level.  Exactly one proton and one $\pi^{-}$, or one
antiproton and one $\pi^{+}$ in the event are required in the
analysis.

Photon candidates are reconstructed by clustering the EMC crystal
energies.  The minimum energy is 25 MeV for barrel showers
($|\cos\theta|<0.80$) and 50 MeV for end-cap showers
($0.86<|\cos\theta|<0.92$). EMC cluster timing requirements are made
to suppress electronic noise and energy deposits unrelated to the
events.

\subsection{\boldmath Selection of $\psip\to\gamma\chicj,\chicj\to\pnpim$}

\begin{figure}[htbp]
\centering
\begin{overpic}[width=8.0cm,height=5.0cm,angle=0]{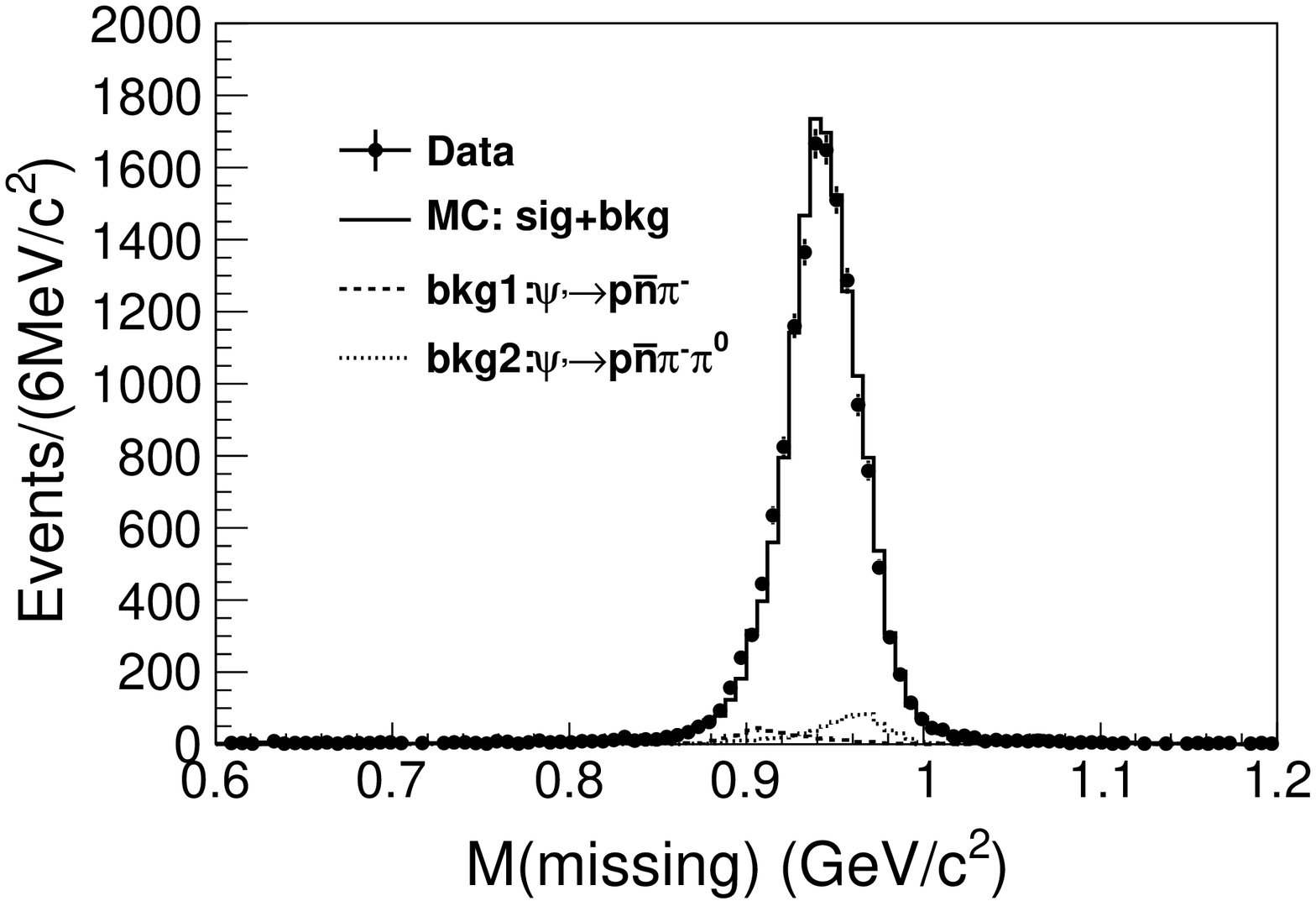}
\put(25,25){\large\bf (a)}
\end{overpic}
\begin{overpic}[width=8.0cm,height=5.0cm,angle=0]{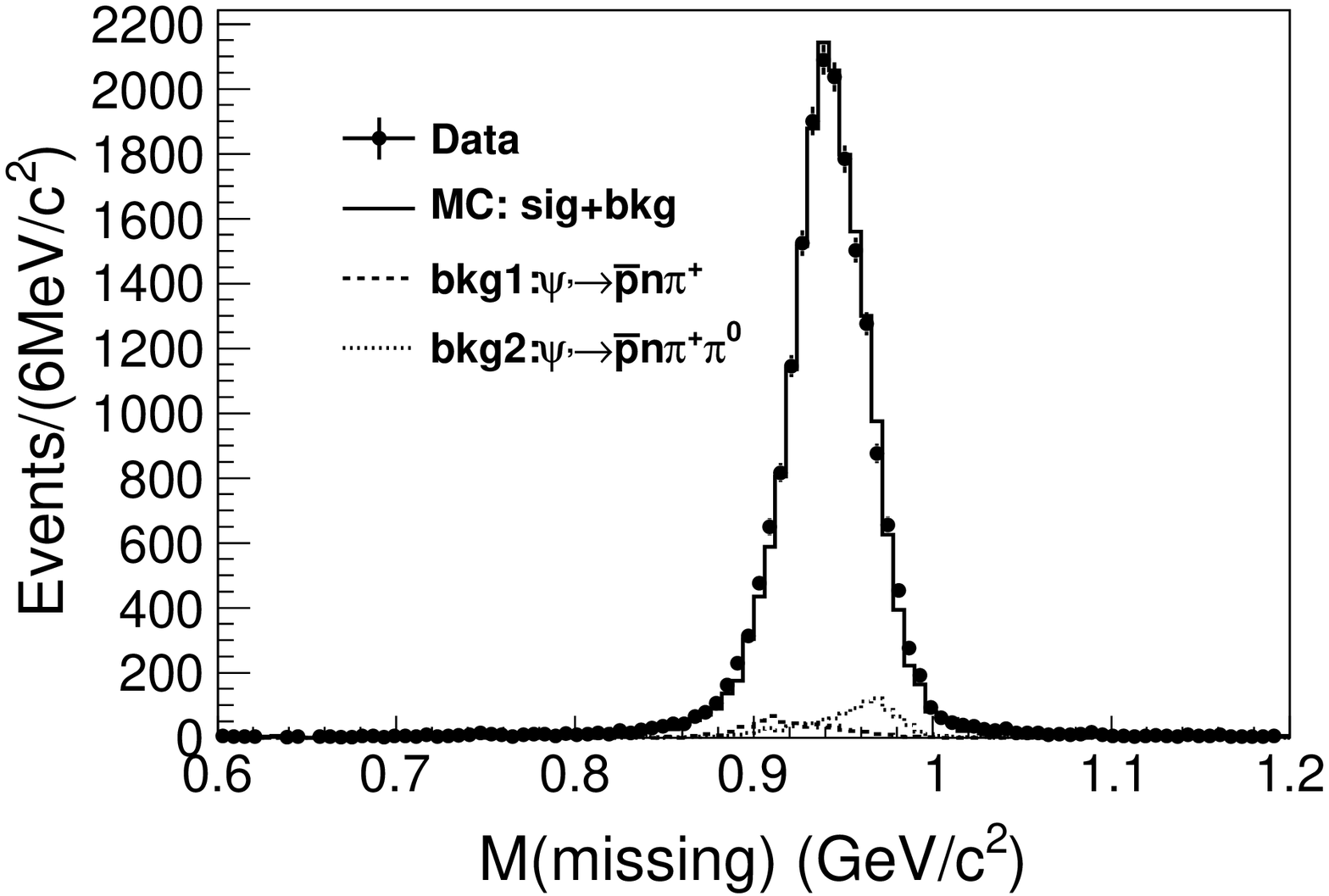}
\put(25,25){\large\bf (b)}
\end{overpic}
\parbox[1cm]{16cm} { \caption{ The distributions of mass recoiling (a)
    against $\gamma p\pi^-$ in $\psip\to\gamma\pnpim$ and (b) against
    $\gamma \bar{p}\pi^+$ in $\psip\to\gamma\pnpip$. Dots with error
    bars are data, and the solid histograms are the sum of signal and
    backgrounds, where the backgrounds are estimated from the
    inclusive MC and the continuum data at $\sqrt{s}=3.65$
    GeV/$c^{2}$.  The dominant background contributions from
    $\psip\to\pnpiccp$ and $\psip\to\pnpipim$ ($\pnpipip$) are shown
    as dashed and dotted lines, respectively.  }
 \label{mass-nbar-n}
}
\end{figure}

For the channel $\psip\to\gamma\chicj,\chicj\to\pnpim$, at least one
photon with energy greater than 80 MeV is required.  To remove photons
coming from interactions of charged particles or neutrons
(antineutrons) in the detector, the angles between the photon and the
antiproton, antineutron and other particles (pion, proton and neutron)
are required to be greater than $30^\circ$, $20^\circ$, and
$10^\circ$, respectively.  A one-constraint (1C) kinematic fit is
performed under the $\psip\to\gamma\pnpim$ hypothesis, where the
neutron (antineutron) is treated as a missing particle.  For events
with more than one isolated photon candidate, the combination with the
smallest $\chi_{1C}^2$ is retained, and $\chi_{1C}^2<10$ is
required. The distributions of the recoiling mass against $\gamma
p\pi^-$ and $\gamma \bar{p}\pi^+$ are shown in Fig.~\ref{mass-nbar-n}
(a) and Fig.~\ref{mass-nbar-n} (b), respectively. Clear neutron and
antineutron peaks are observed around 0.938 GeV/$c^2$.

An antineutron can form a cluster in the EMC with very high
probability due to annihilation in the detector. To further purify the
events with antineutrons and more than one photon, $\alpha<15^{\circ}$
is required, where $\alpha$ is the angle between the expected
antineutron direction and the nearest photon. Further,
$M_{p\pi^{-}(\bar{p}\pi^{+})}>$ 1.15 GeV/$c^2$ is required to remove
background events with $\Lambda$ or $\bar{\Lambda}$. Finally, the
transverse momentum for the proton or antiproton is required to be
greater than 0.3 GeV/$c$ due to the difference of the tracking
efficiency at low transverse momentum between data and MC.

\subsection{\boldmath Selection of $\psip\to\gamma\chicj,\chicj\to\pnpipim$}

\begin{figure}[htbp]
\centering
\begin{overpic}[width=8.0cm,height=5.0cm,angle=0]{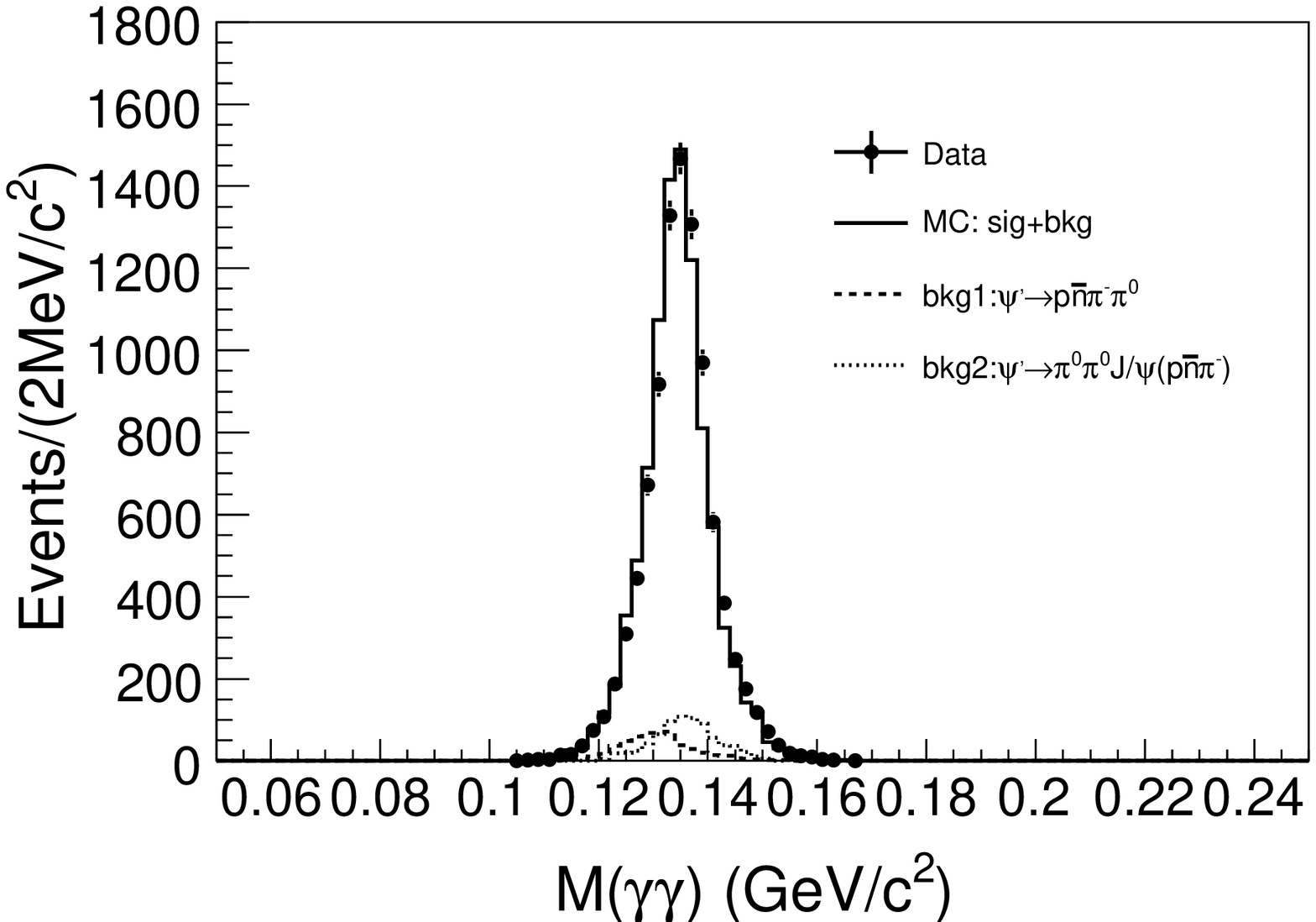}
\put(22,25){\large\bf (a)}
\end{overpic}
\begin{overpic}[width=8.0cm,height=5.0cm,angle=0]{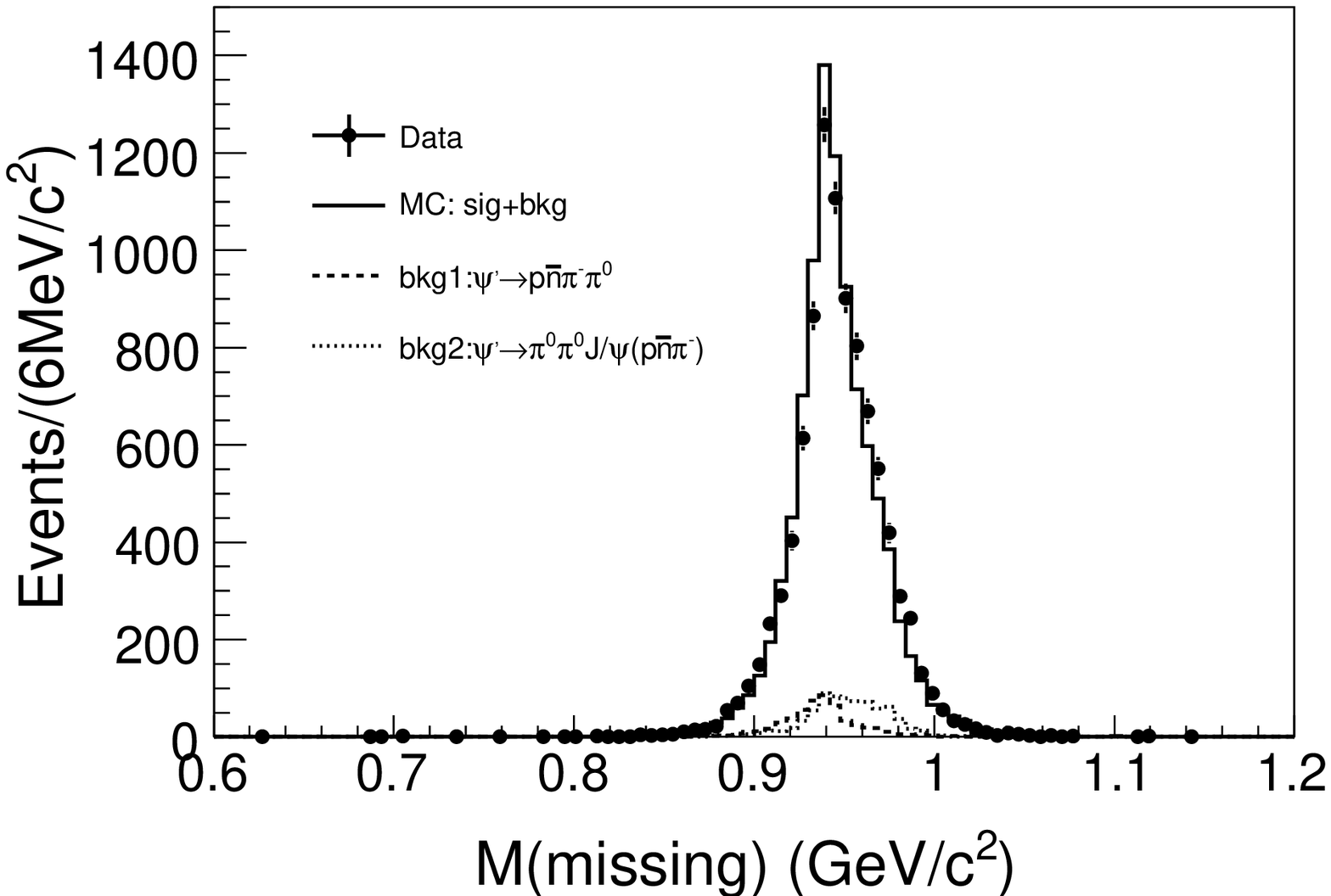}
\put(22,25){\large\bf (b)}
\end{overpic}
\parbox[1cm]{16cm} { \caption{ The invariant mass distribution of (a)
    $\gamma\gamma$ from $\pi^{0}$ and (b) the distribution of
    recoiling mass against $\gamma p\pi^{-}\pi^0$ in
    $\psip\to\gamma\pnpipim$. Dots with error bars are data, and
    the solid histograms show the sum of signal and backgrounds, where
    the backgrounds are estimated from the inclusive MC and the
    continuum data at $\sqrt{s}=3.65$ GeV/$c^{2}$. The dominant
    background contributions from $\psi(2S)\to p\bar{n}\pi^{-}\pi^{0}$
    and $\psi(2S)\to\pi^{0}\pi^{0}J/\psi$,$J/\psi\to p\bar{n}\pi^{-}$
    are shown as dashed and dotted lines, respectively.  }
  \label{mass-pi0-nbar}
}
\end{figure}

For $\psip\to\gamma\chicj,\chicj\to\pnpipim$, at least three isolated
photons are required, where the photon isolation criteria are the same
as those in $\psip\to\gamma\chicj,\chicj\to\pnpim$.  $\pi^{0}$
candidates are selected from any pair of photon candidates that can be
kinematically fitted to the $\pi^0$ mass and satisfy $\chi^{2}<20$.
There must be at least one $\pi^{0}$ candidate.  A 1C kinematic fit is
performed to the $\psip\to\threegamma\pnpim$ hypothesis constraining
the mass of the missing particle to that of the neutron, where the
three photons are a $\pi^0$ candidate together with another
photon. Kinematic fits are carried out over all possible three-photon
combinations.  The kinematic fit with the smallest
$\chi^2_{\threegamma\pnpim}$ is retained, and
$\chi^2_{\threegamma\pnpim}<10$ is required. If there is more than one
$\pi^0$ candidate among the three photons, the pair with the invariant
mass closest to the $\pi^{0}$ mass is assigned to be from the
$\pi^{0}$ decay.  To further suppress backgrounds with final states
$\twogamma\pnpim$ and $\fourgamma\pnpim$, 1C kinematic fits are
performed under $\twogamma\pnpim$ and $\fourgamma\pnpim$
hypotheses,\footnote{The two-photon combinations are pairs of photons
  from $\pi^0$ candidates, and the four-photon combination are pairs
  of photons from $\pi^0$ candidates together with another two
  photons. If there are multiple combinations, the one with the
  smallest $\chi^2$ is selected.}  and
$\chi^2_{\threegamma\pnpim}<\chi^2_{\twogamma\pnpim}$ and
$\chi^2_{\threegamma\pnpim}<\chi^2_{\fourgamma\pnpim}$ are required.
With the above selection criteria, the invariant mass of $\pi^0$ and
the distribution of the recoiling mass against $\gamma p\pi^{-}\pi^0$
in $\psip\to\gamma\pnpipim$ are shown in Fig.~\ref{mass-pi0-nbar} (a)
and Fig.~\ref{mass-pi0-nbar} (b), respectively. Similar distributions
are obtained in the charge conjugate channel.

To suppress the background from $\psip\to\pipizero\jpsi$, for events
with at least four photons, $\pipizero$ combinations are formed with
any four photons, the one with the smallest
$\Delta=\sqrt{(m_{\gamma1\gamma2}-m_{\pi^0})^2+(m_{\gamma3\gamma4}-m_{\pi^0})^2}$
is selected, and $|M_{recoil}-M_{\jpsi}|>$ 50 MeV/$c^2$ is required,
where $M_{recoil}$ is the mass recoiling against $\pipizero$.  To
purify the channel with antineutrons, the same requirement
$\alpha<15^\circ$ as that for the channel
$\psip\to\gamma\chicj,\chicj\to\pnpim$ is applied, and the transverse
momentum for proton or antiproton is required to be greater than 0.3
GeV/$c$ in order to reduce the systematic uncertainty.

\begin{figure}[htbp]
\centering
\begin{overpic}[width=8.0cm,height=5.0cm,angle=0]{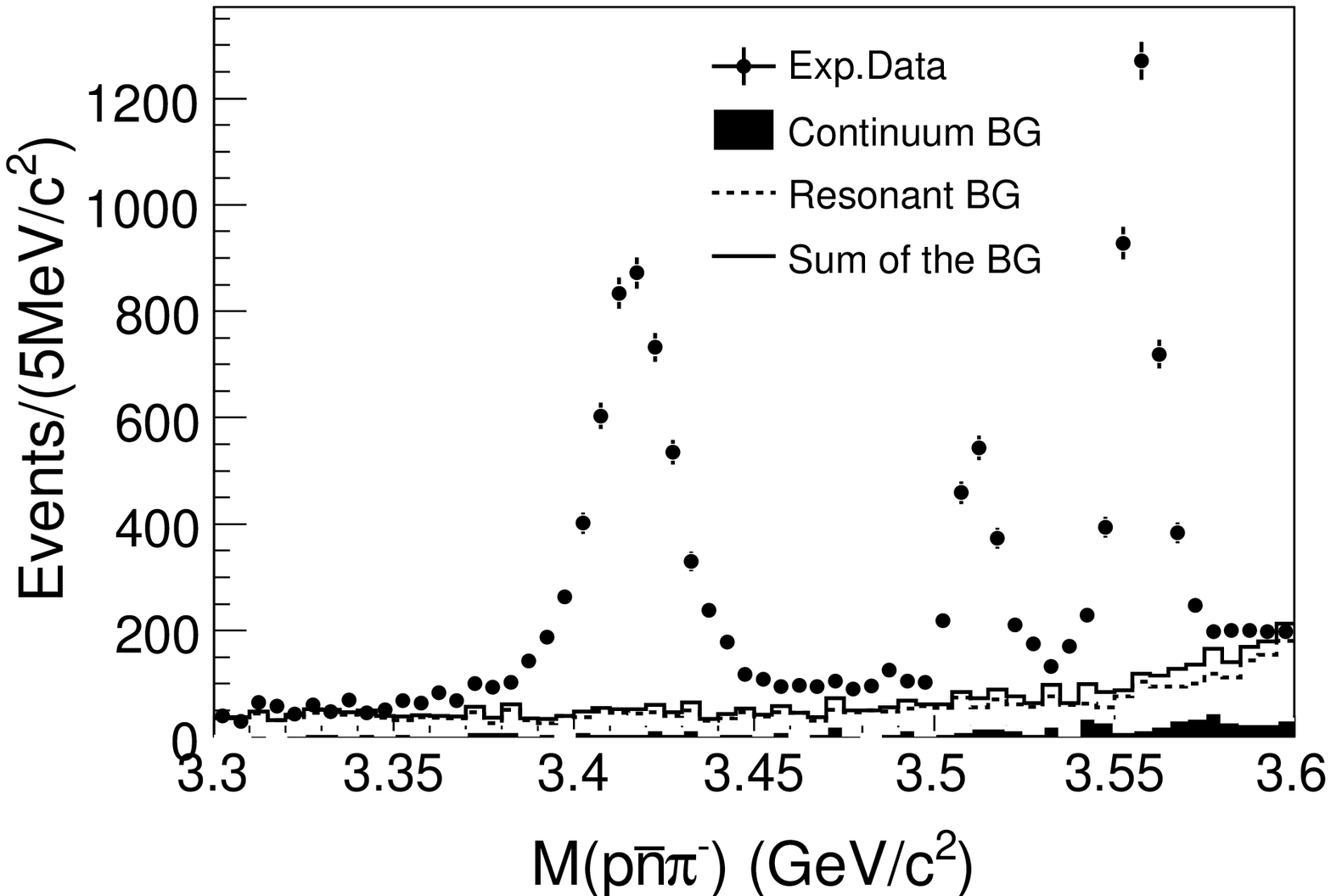}
\put(22,30){\large\bf (a)}
\end{overpic}
\begin{overpic}[width=8.0cm,height=5.0cm,angle=0]{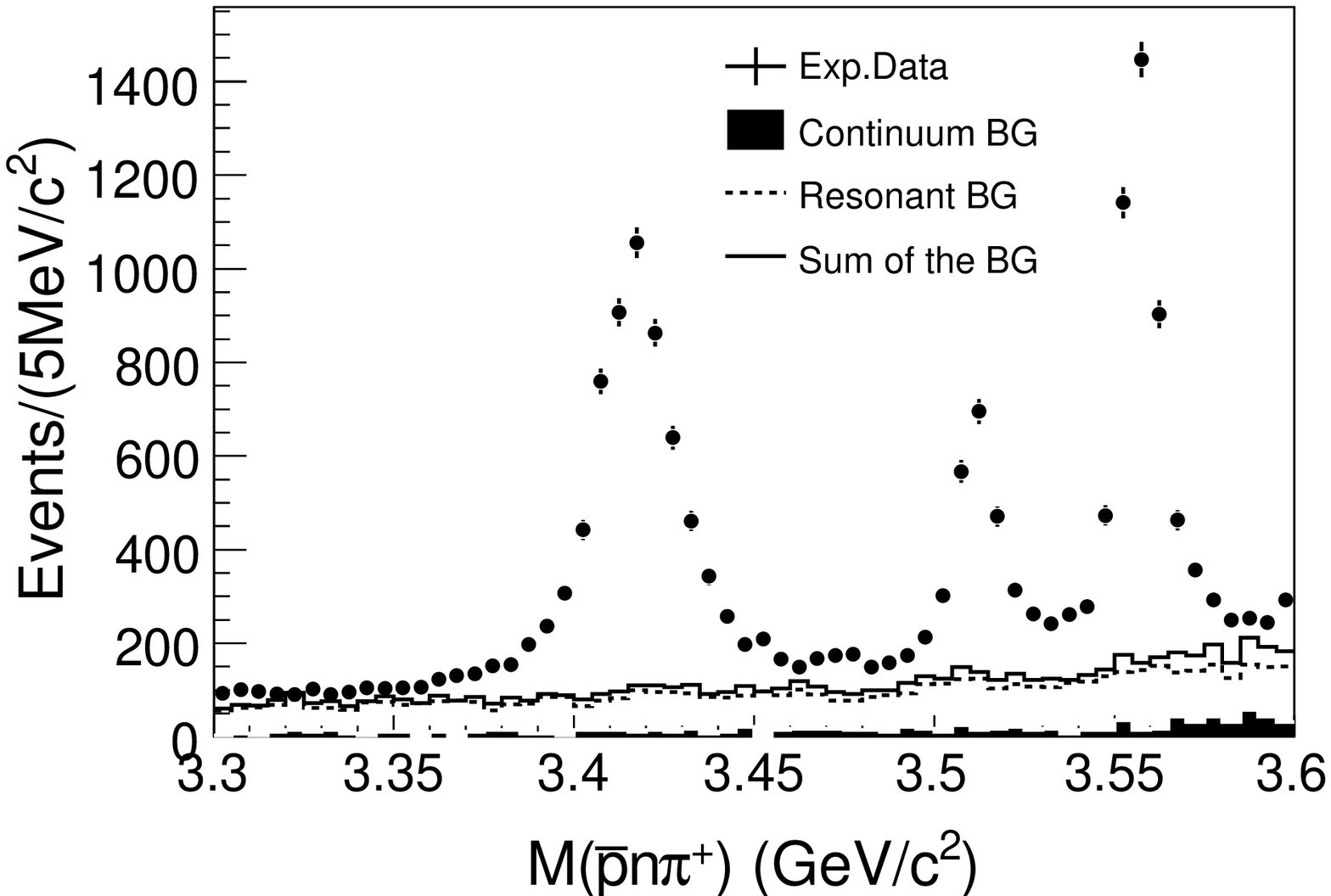}
\put(22,30){\large\bf (b)}
\end{overpic}
\parbox[1cm]{16cm} {
 \caption{
  Invariant mass distribution of (a) $\pnpim$ for $\psip \to \gamma\pnpim$
  events, and of (b) $\pnpip$ for the charge conjugate channel.
  Dots with error bars are data, and the filled histogram is
   the normalized non-resonant contribution estimated from continuum data
  at $\sqrt{s}=3.65$ GeV/$c^{2}$. Resonant background, shown as the dashed
histogram, is dominant and is estimated from MC.
The sum of both background contributions is shown as the solid histogram.
  \label{mass-pnpi}
 }
}

\end{figure}

\begin{figure}[htbp]
\centering
\begin{overpic}[width=8.0cm,height=5.0cm,angle=0]{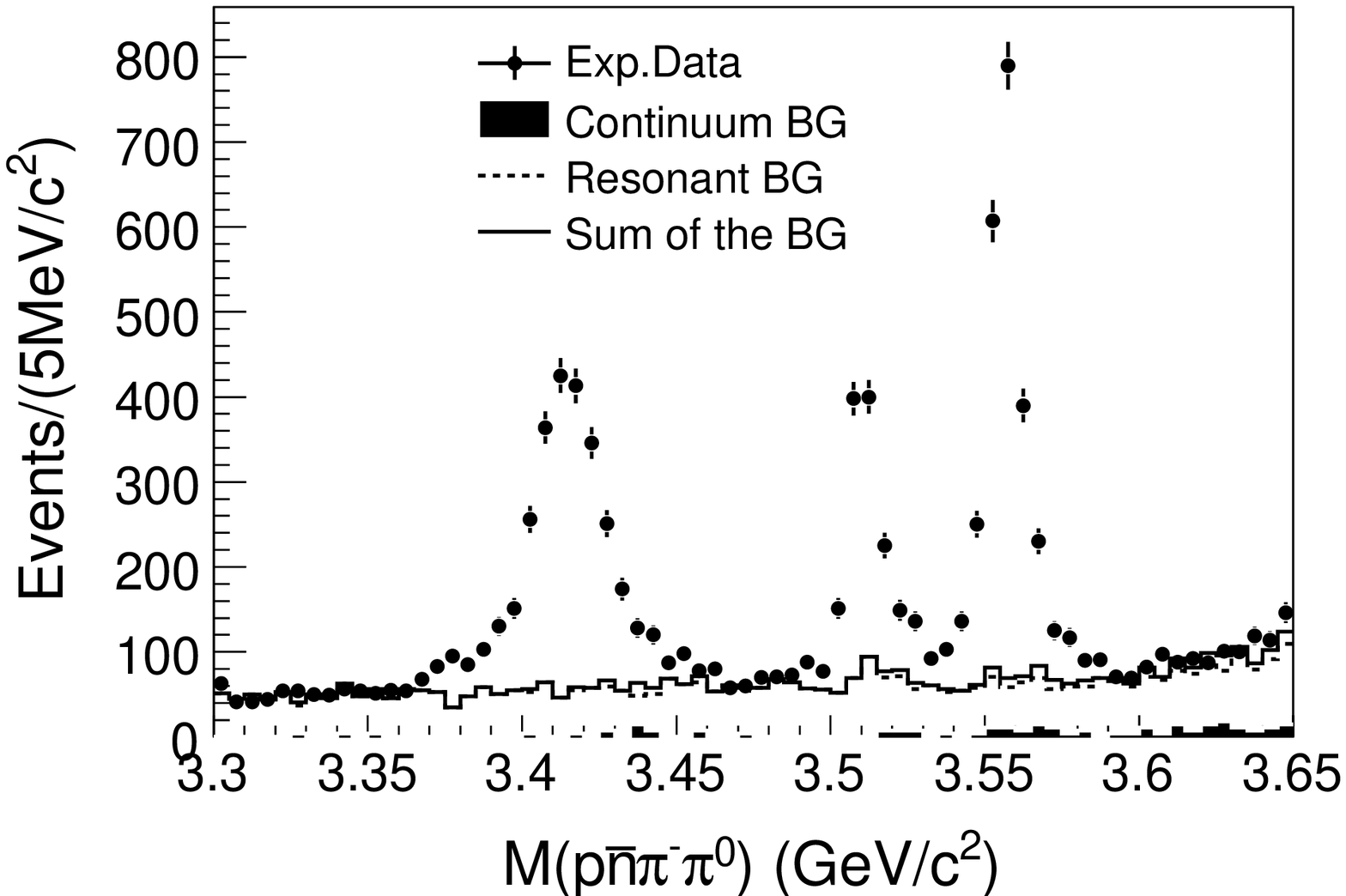}
\put(22,30){\large\bf (a)}
\end{overpic}
\begin{overpic}[width=8.0cm,height=5.0cm,angle=0]{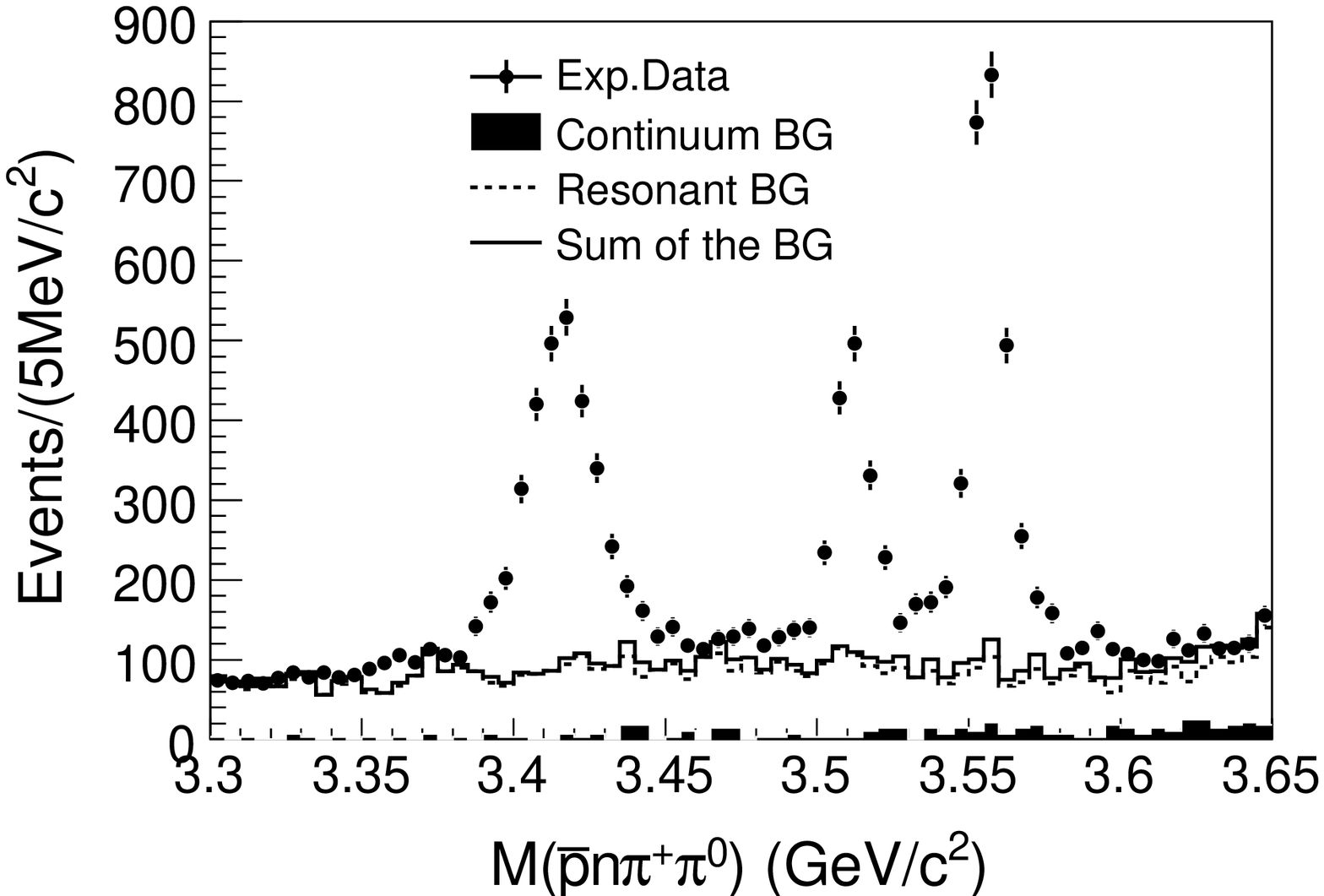}
\put(22,30){\large\bf (b)}
\end{overpic}
\parbox[1cm]{16cm} {
 \caption{
   Invariant mass distribution of (a) $\pnpipim$ for the $\psip \to \gamma\pnpipim $
   channel, and of (b) $\pnpipip$ for the charge conjugate channel.
   Dots with error bars are data, and the filled histogram is the
   normalized non-resonant background
  contribution estimated from continuum data at $\sqrt{s}=3.65$ GeV/$c^{2}$.
Resonant background, shown as the dashed
histogram, is dominant and is estimated from MC.
The sum of both background contributions is shown as the solid histogram.
   \label{mass-pnpipi0}
 }
}
\end{figure}

With the above criteria, clear $\chicj$ signals are observed in the
invariant mass distributions of $\pnpim$($\pnpip$) in
$\chicj\to\pnpim(\pnpip)$ and of $\pnpipim$($\pnpipip$) in
$\chicj\to\pnpipim(\pnpipip)$, as shown in Figs.~\ref{mass-pnpi} (a)
and (b) and Figs.~\ref{mass-pnpipi0} (a) and (b), respectively.

\section{Background studies}
Background is investigated using both exclusive and inclusive MC samples,
 as well as continuum data. For $\psip\to\gamma\chicj,\chicj\to\pnpim$,
 a number of individual channels from the inclusive MC sample have
 been found to contribute to the background. The following reactions
 have been found to be important: $\psip\to\pnpipim$ ($\approx$ 20\%
 of the total background), $\psip\to\pnpim$ (8\%),
 $\psip\to\pipizero\jpsi$ with $\jpsi\to\pnpim$ (1\%), and
 $\psip\to\gamma\chicj$ with $\chicj\to\pnpipim$ (4\%). For the first
 three background channels, the world average branching fractions
 listed in PDG~\cite{pdg} are used for normalization, and for
 $\chicj\to\pnpipim$, we use our own measurement as described in
 Sect.~\ref{sect:bf}.  The background originating from non-resonant
 processes is estimated using a continuum data sample collected at a
 center-of-mass energy of 3.65 GeV/$c^{2}$ after normalizing it to the
 integrated luminosity and the production cross section. The invariant
 mass distributions of $\pnpim$ and of the charged conjugate state
 $\pnpip$ are shown in Figs.~\ref{mass-pnpi} (a) and (b),
 respectively. No peaking background is observed in the signal region.

 Also $\psip\to\gamma\chicj,\chicj\to\pnpipim$ contains a number of
 individual background channels according to studies with the
 inclusive MC sample.  The dominant sources are the reactions
 $\psip\to\pnpipim$ ($\approx$ 25\% of the total background),
 $\psip\to\pipizero\jpsi$ with $\jpsi\to\pnpim$ (10\%), and
 $\psip\to\gamma\chicj$ with $\chicj\to\pnpim$ (1\%).  Again, for the
 latter channel the branching fraction obtained from our own
 measurement is used for normalization (see Sect.~\ref{sect:bf}); for
 the first two background channels the PDG~\cite{pdg} branching
 fractions are used. Additional background is studied using the
 inclusive MC sample, and the continuum data sample.  The invariant
 mass distributions of $\pnpipim$ and $\pnpipip$ are shown in
 Fig.~\ref{mass-pnpipi0} (a) and Fig.~\ref{mass-pnpipi0} (b),
 respectively.  Also in this case, no peaking background is observed
 in the signal region.  It has been explicitly verified that the
 process $\psip\to\gamma\chicj,\chicj\to\gamma \jpsi$ with
 $\jpsi\to\pnpim$ or $\jpsi\to\pnpipim$ does not represent a
 significant source of background.

\section{Intermediate states}
\label{sect:struct}
\begin{figure}[htbp]
\centering
\begin{overpic}[width=6.0cm,height=4.0cm,angle=0]{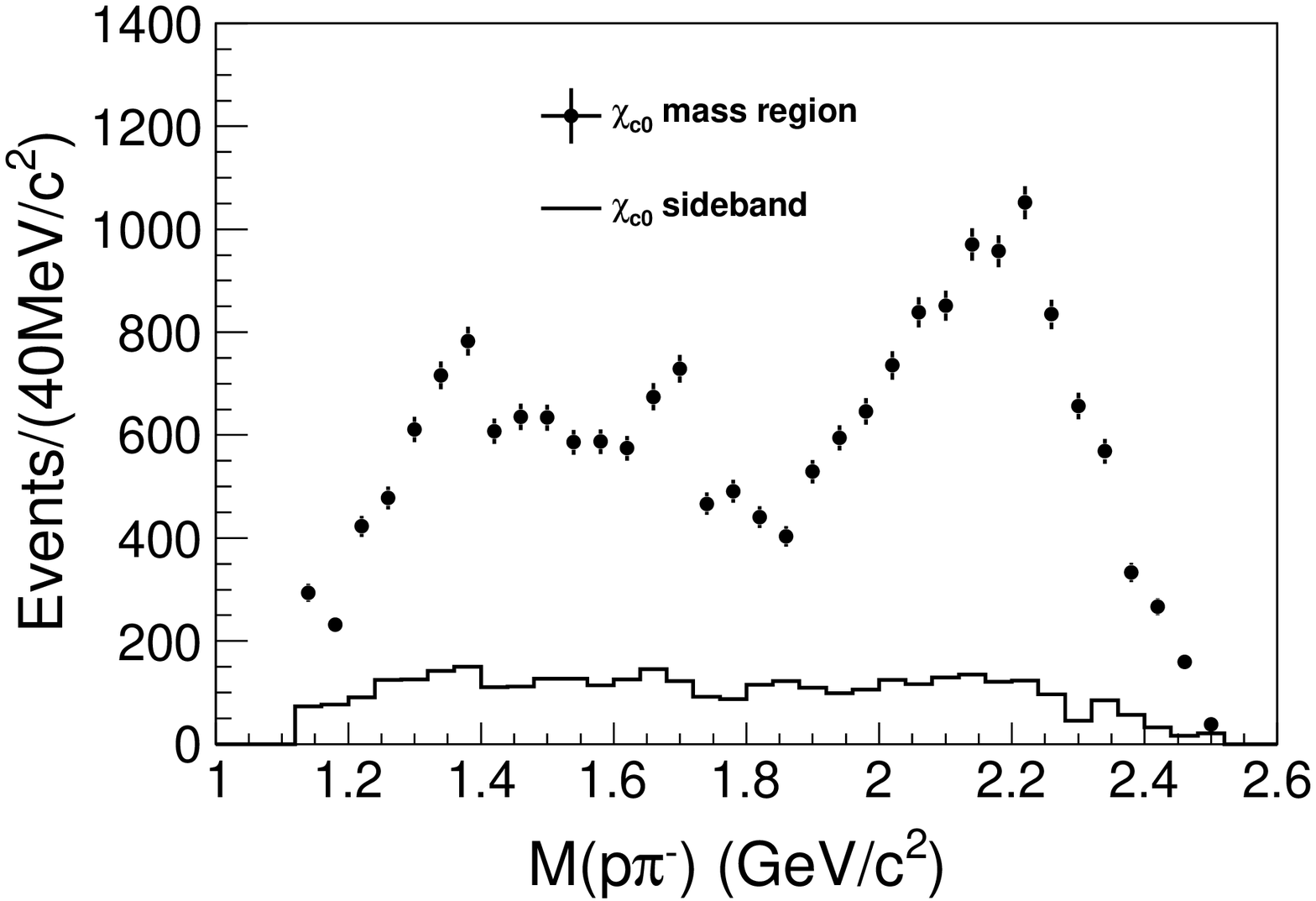}
\put(20,50){\small\bf (a)}
\end{overpic}
\begin{overpic}[width=6.0cm,height=4.0cm,angle=0]{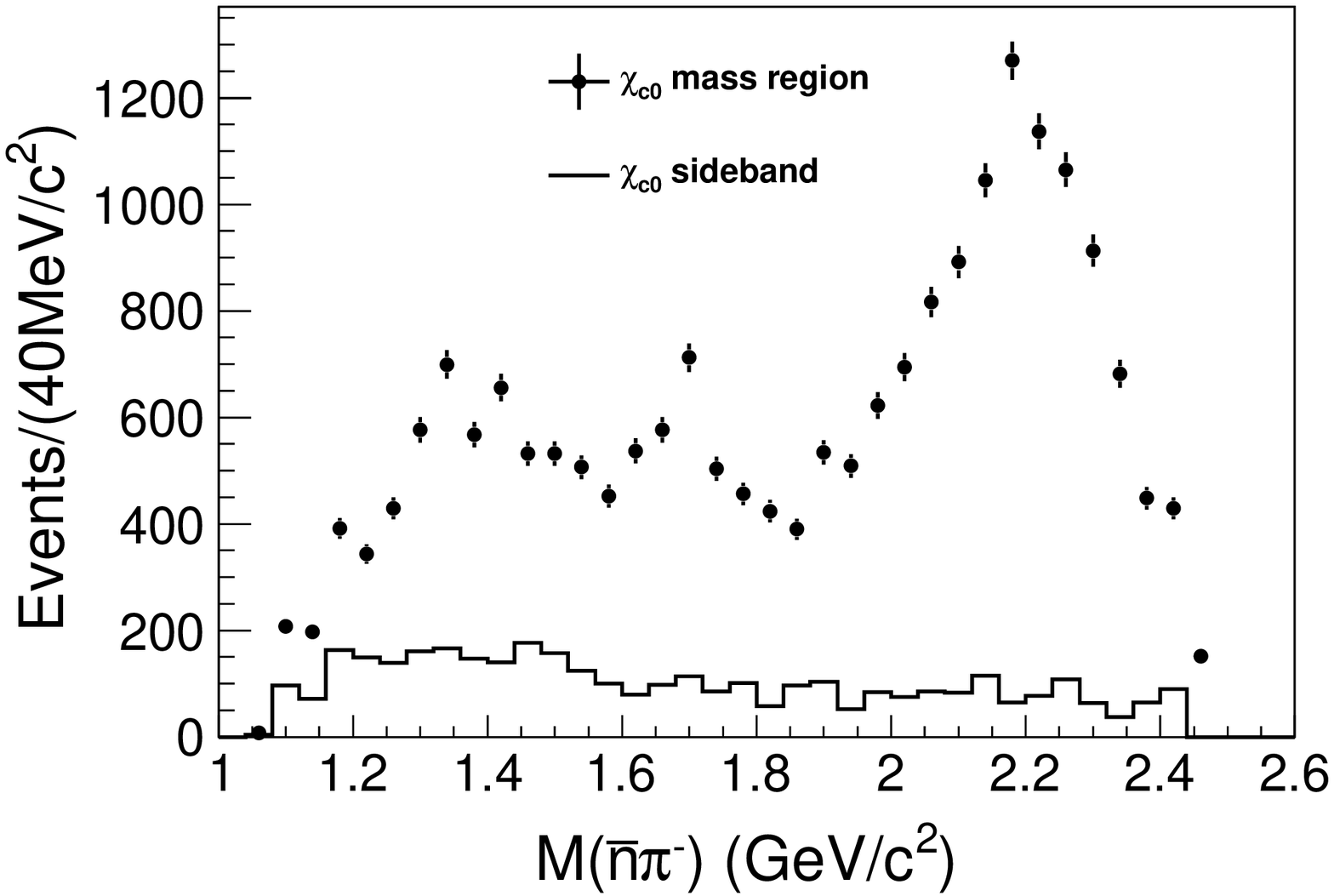}
\put(20,50){\small\bf (b)}
\end{overpic}
\begin{overpic}[width=6.0cm,height=4.0cm,angle=0]{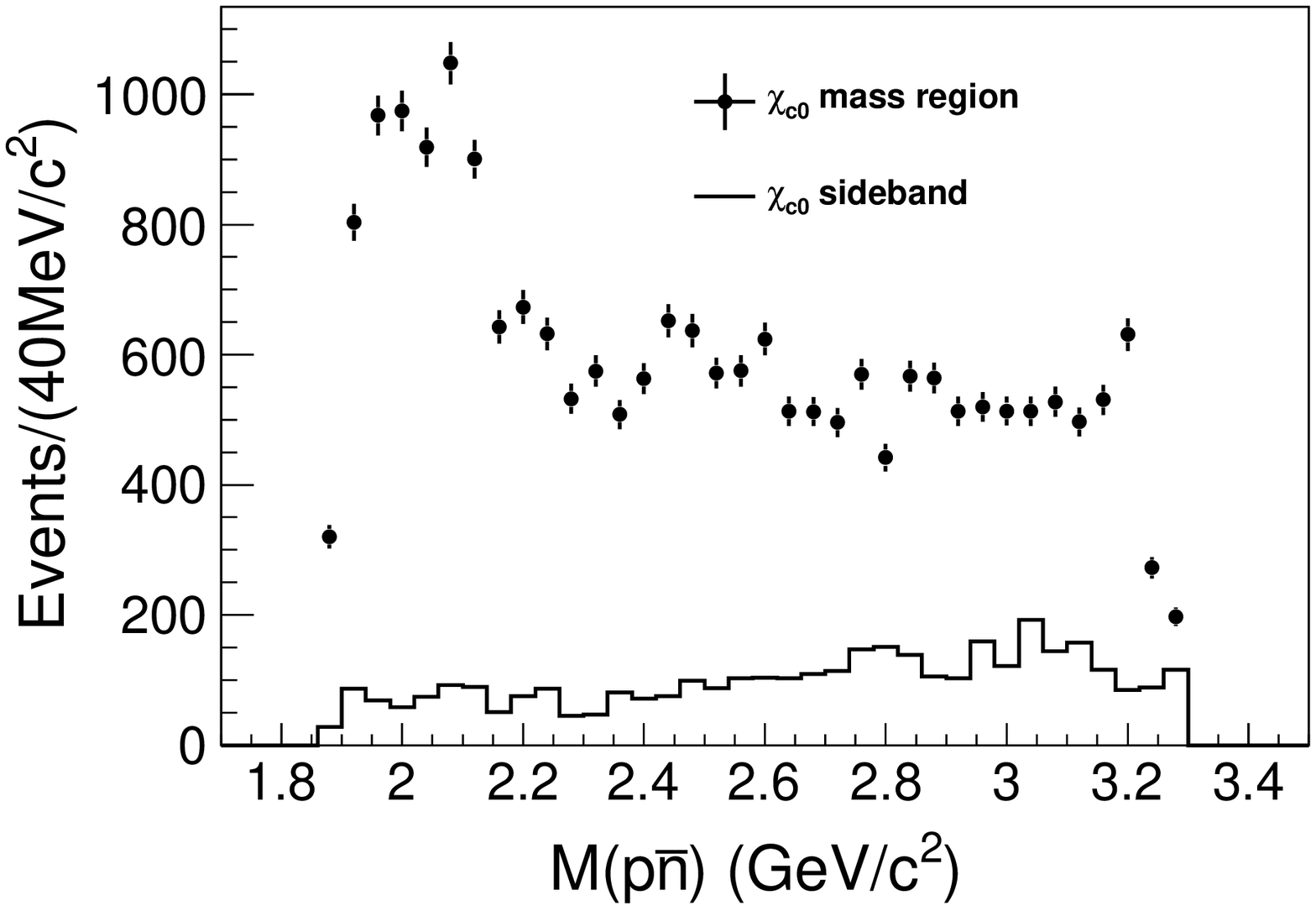}
\put(20,50){\small\bf (c)}
\end{overpic}
\begin{overpic}[width=6.0cm,height=4.0cm,angle=0]{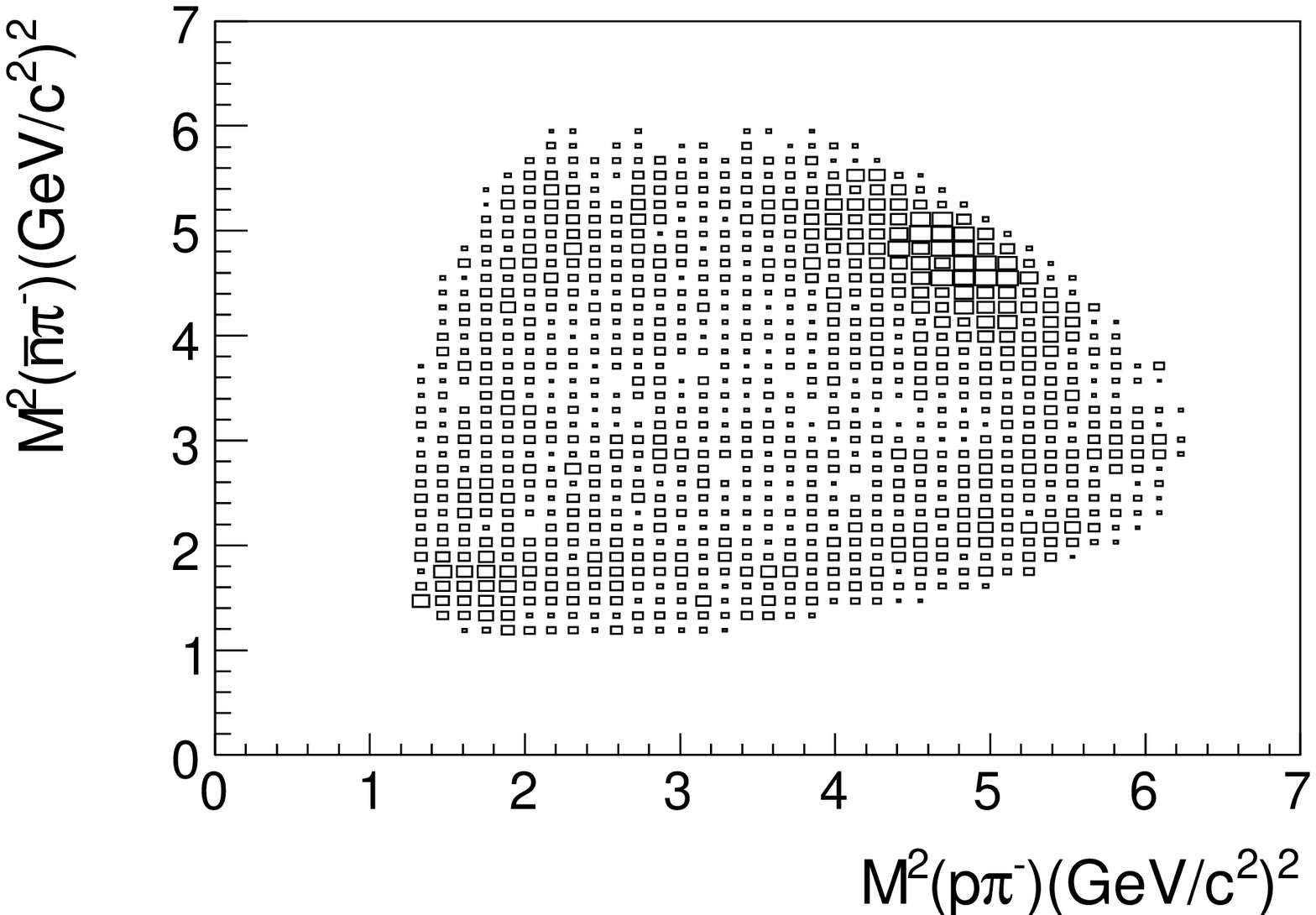}
\put(20,50){\small\bf (d)}
\end{overpic}
\parbox[1cm]{15cm} {
  \caption{
    The invariant mass distributions with the efficiency correction of (a) $p\pi^-$, (b) $\bar{n}\pi^-$, (c)
    $p\bar{n}$ and (d) Dalitz plot for $\chiczero\to\pnpim$ events.
    The dots with error bars are data, and the histograms are for backgrounds
    obtained from sideband events.
    \label{Nstar-plot}
  }
}
\end{figure}

We have searched for potential intermediate $N^\ast$ baryon resonances
decaying into either $p\pi^-$ or $\bar{n}\pi^-$. Such a study is
needed for correct modeling of $\chicj\to\pnpim$ and
$\chicj\to\pnpipim$ in order to determine the efficiencies used in the
calculation of the branching fractions of these decays, which is the
main purpose of this work.  Selecting $\chiczero$ events in the
$\pnpim$ decay mode, $|M_{\pnpim}-M_{\chiczero}|<$ 45 MeV/$c^2$, the
efficiency corrected $p\pi^-$ and $\bar{n}\pi^-$ invariant mass
distributions are shown in Fig.~\ref{Nstar-plot} (a) and
Fig.~\ref{Nstar-plot} (b), respectively. Background contributions are
also shown and are obtained from the sideband invariant mass region 45
MeV/$c^2$ $<|M_{\pnpim}-M_{\chiczero}|<$ 75
MeV/$c^2$. Fig.~\ref{Nstar-plot} (d) shows in addition the
corresponding Dalitz plot $M^2(\bar{n}\pi^-)$ vs. $M^2(p\pi^-)$. The
structures of the $N^\ast$ states at around 1.4 GeV/$c^2$ and 1.7
GeV/$c^2$ can be seen in both the $p\pi^-$ and $\bar{n}\pi^-$
invariant mass spectra as well as the bands in the Dalitz plot.
The $p\bar{n}$ invariant mass is shown in Fig.~\ref{Nstar-plot} (c).
A large enhancement in the $p\bar{n}$ threshold region is observed,
which is also visible as a diagonal band along the upper right-band
edge in the Dalitz plot in Fig.~\ref{Nstar-plot} (d).  A similar
threshold enhancement has qualitatively been observed elsewhere, such
as a $p\bar{p}$ threshold enhancement in B meson
decays~\cite{belleb2ppk,belleb2ppd, babarb2ppk,babarb2ppd,
  belleb2ppkp,belleb2ppkkp}, $\psip$ decays~\cite{cleopsipppp}, and in
the shape of the timelike electromagnetic form factor of the proton
measured at BaBar~\cite{babartlff}.

The peak at around 2.2 GeV/$c^2$ in both the $p\pi^-$ and $\bar{n}\pi^-$
invariant mass spectra is partly due to the reflection of the threshold
enhancement of $p\bar{n}$. It might also be partly due to high mass $N^\ast$
states, such as $N^{\ast}(2190)$ or $N^{\ast}(2220)$.
The same structures are observed in the charge conjugate mode of $\chiczero\to\pnpip$
and the decays of $\chi_{c1,2}\to\pnpim$. A partial wave analysis is
necessary to obtain more information about the $N^\ast$ components and
the threshold enhancement in the invariant mass distribution of $p\bar{n}$.

\begin{figure}[htbp]
\centering
\hskip -0.4cm \mbox{
\begin{overpic}[width=8.0cm,height=5.0cm,angle=0]{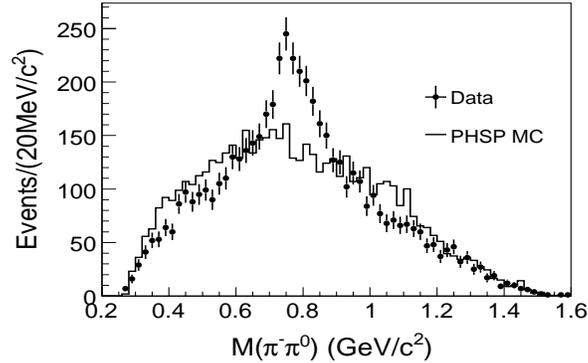}
\put(25,40){{\LARGE }}
\end{overpic}}
\hskip 0.5cm
\parbox[1cm]{16cm}{ \caption{ The invariant $\pi^{-}\pi^{0}$ mass
    distribution in $\chi_{cJ}\to p\bar{n}\pi^{-}\pi^{0}$ where a
    significant $\rho$ signal is observed in the data.}
 \label{rho}
}
\end{figure}

In $\chicj\to\pnpipim$, various two-body and three-body invariant mass
distributions for events within the $\chicj$ signal region were
investigated as well.  No obvious $N^{\ast}$ state is observed. The
distributions are very similar to those of phase space, except the
$\pi^{\pm}\pi^0$ invariant mass spectra that show a significant
$\rho^{\pm}$ signal (see Fig.~\ref{rho}).

\section{Signal extraction}
\label{sect:bf}
Signal yields are extracted using unbinned maximum likelihood fits to
the observed $\pnpim$ and the $\pnpipim$ invariant mass distributions.
The following formula has been used for the fit:
\begin {equation}
  \sum_{i=0}^{2}BW(m; M_{i}; \Gamma_{i})\otimes G(m; \sigma_{i})+BG,
\end {equation}
where $BW(m; M_{i}$; $\Gamma_{i}$) is the Breit-Wigner function for
the natural lineshape of the $\chicj$ resonance, BG represents the
background shape and is described by a third order Chebychev
polynomial, and G(m; $\sigma_{i}$) is a modified Gaussian function
parameterizing the instrumental mass resolution, which was used
by ZEUS Collaboration in ref~\cite{modGaussian} and expressed by:
\begin {equation}
  G(m; \sigma_{i})=\frac {1}{\sqrt{2\pi}\sigma_{i}}
  e^{-(|\frac{m}{\sigma_{i}}|)^{1+(\frac{1}{1+|\frac{m}{\sigma_{i}}|})}}.
\end {equation}
In the fit, the natural widths of the $\chi_{cJ}$ states are fixed to
the PDG~\cite{pdg} values, while their masses and corresponding
instrumental resolutions are floated.  For
$\psip\to\gamma\chicj,\chicj\to\pnpim$, this fit is performed in the
mass region of 3.30 GeV/$c^{2} \le M(\pnpim) \le$ 3.60 GeV/$c^{2}$,
for the process $\psip\to\gamma\chicj,\chicj\to\pnpipim$ the mass
region of 3.30 GeV/$c^{2}\le$ 3.64 GeV/$c^{2}$ has been chosen.

\begin{figure}[htbp]
\centering
\begin{overpic}[width=8.0cm,height=5.0cm,angle=0]{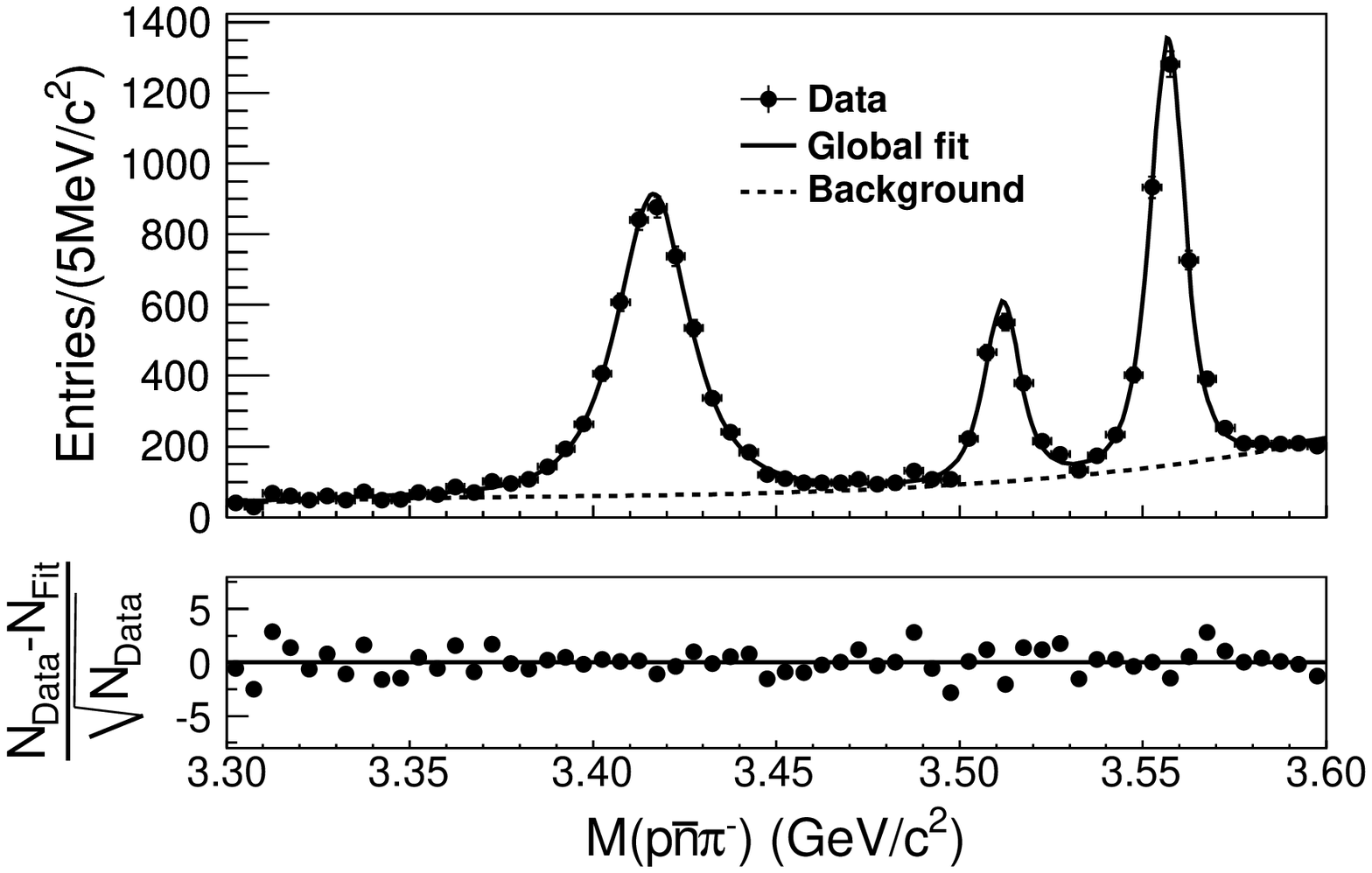}
\put(25,40){\large\bf(a)}
\end{overpic}
\begin{overpic}[width=8.0cm,height=5.0cm,angle=0]{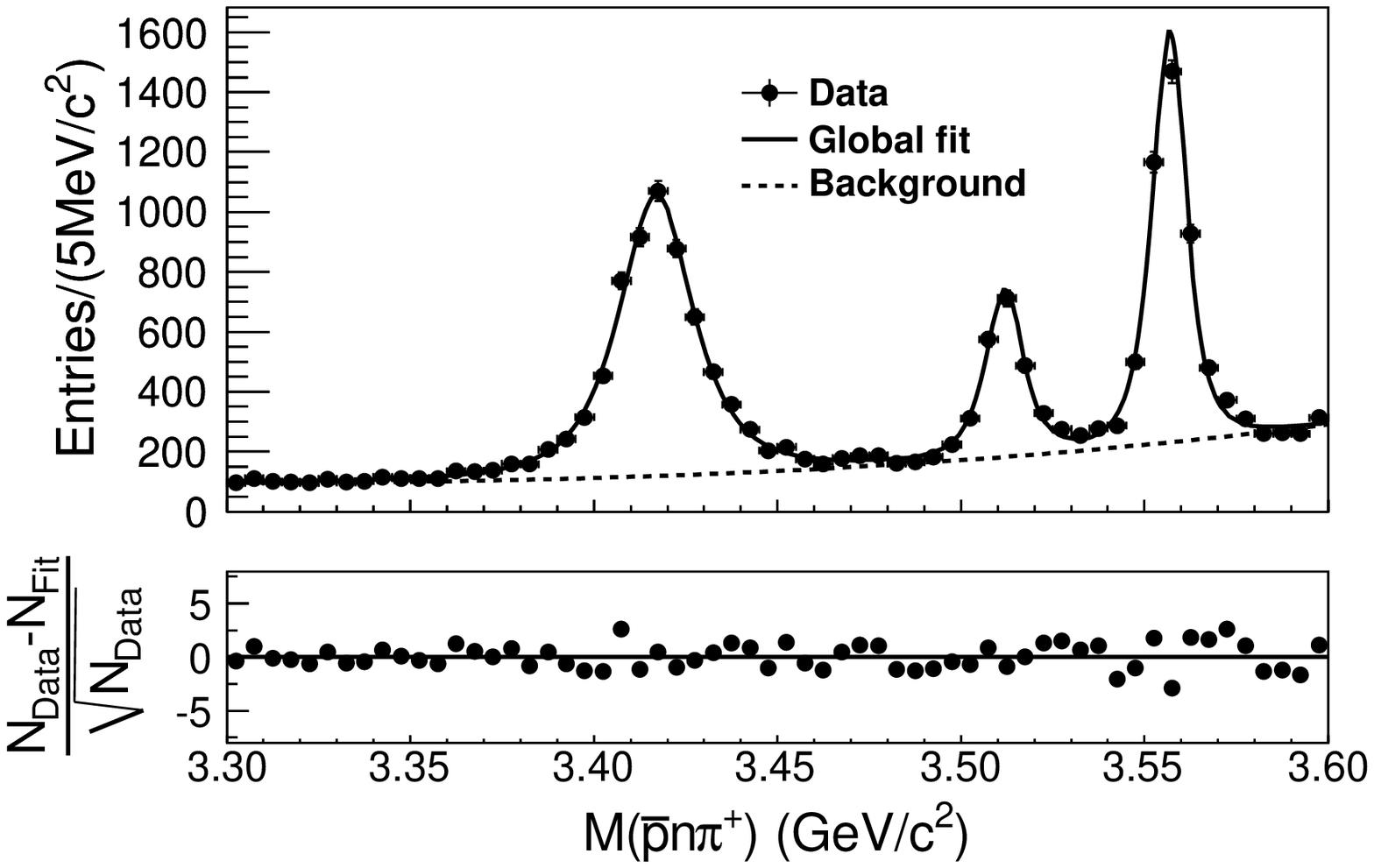}
\put(25,40){\large\bf(b) }
\end{overpic}
\parbox[1cm]{16cm} { \caption{ Upper plot: The fit to the invariant
    mass distributions of (a) $\pnpim$ and (b) the charge conjugate
    state $\pnpip$.  Dots with error bars are data, the solid curve is
    showing the fit to signal events, and the dashed line is the
    fitted background distribution. Lower plot: The distribution of
    $\frac{N_{Data}-N_{Fit}}{\sqrt{N_{Data}}}$ from the
    fit.  \label{fitpnpi} }
}
\end{figure}

\begin{figure}[htbp]
\centering
\begin{overpic}[width=8.0cm,height=5.0cm,angle=0]{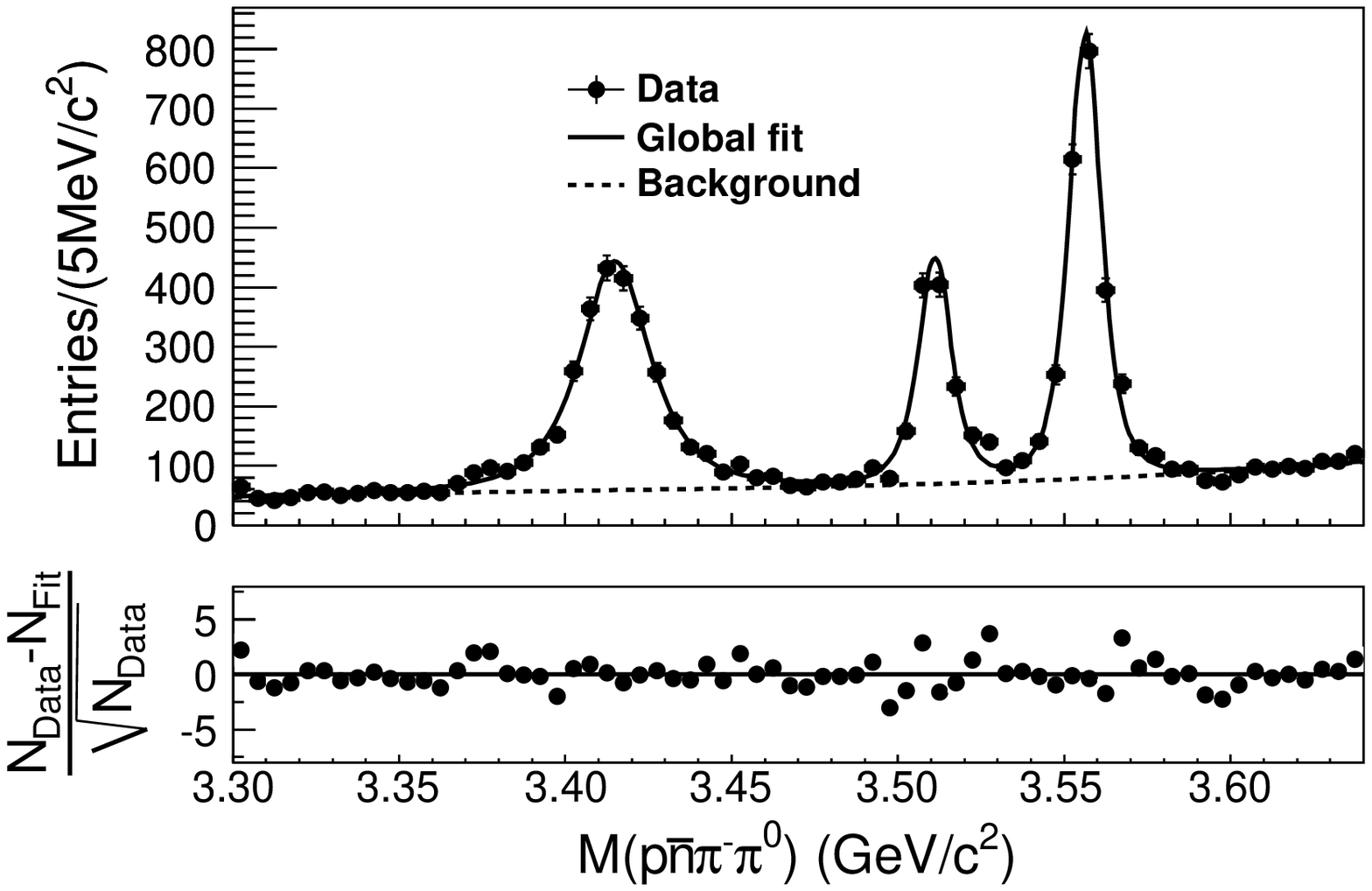}
\put(25,40){\large\bf (a)}
\end{overpic}
\begin{overpic}[width=8.0cm,height=5.0cm,angle=0]{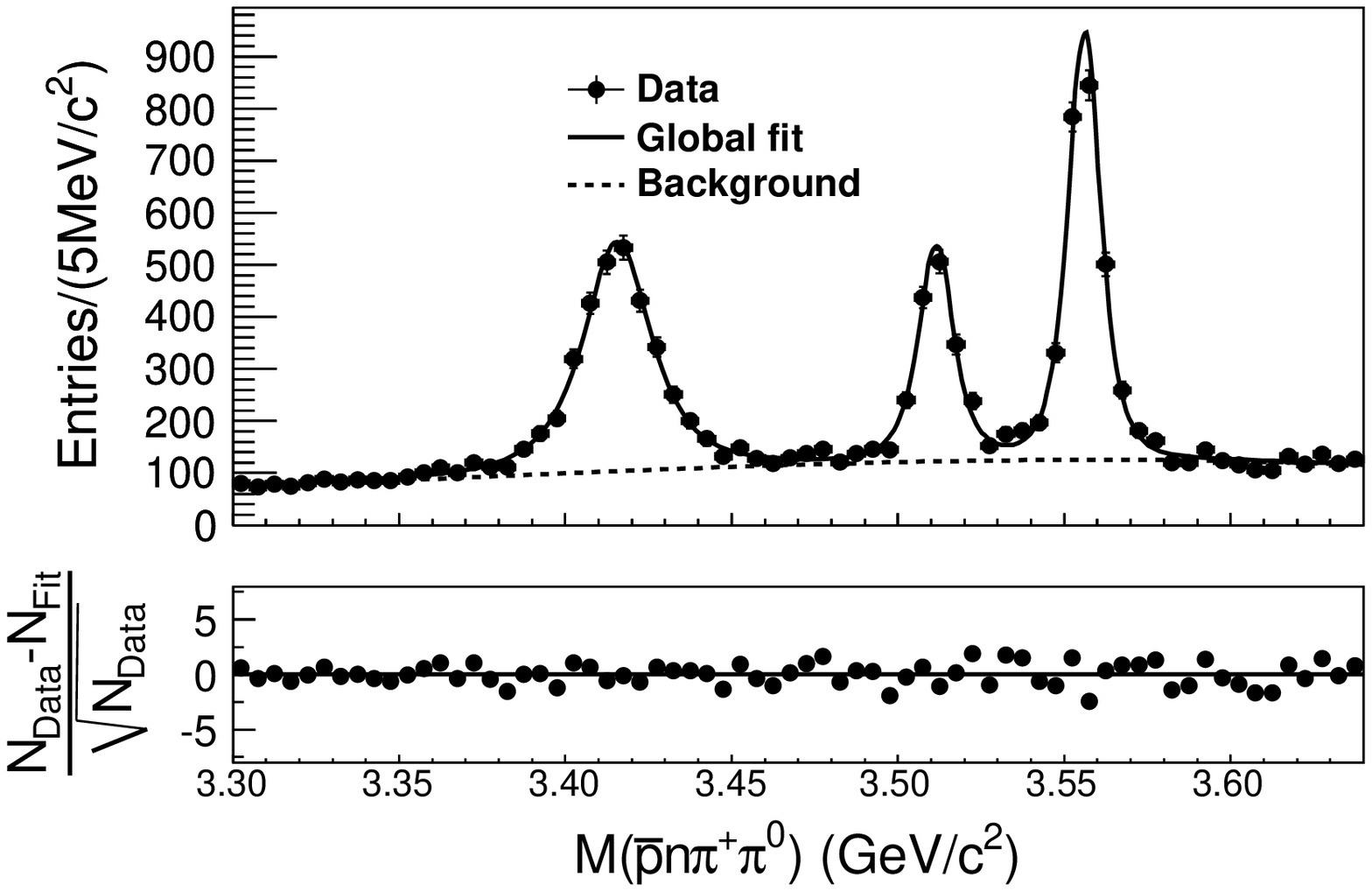}
\put(25,40){\large\bf(b)}
\end{overpic}
\parbox[1cm]{16cm} { \caption{ Upper plot: The fit to the invariant
    mass distributions of (a) $\pnpipim$ and (b) the charge conjugate
    state $\pnpipip$.  Dots with error bars are data, the solid curve
    is showing the fit to signal events, and the dashed line is the
    fitted background distribution. Lower plot: The distribution of
    $\frac{N_{Data}-N_{Fit}}{\sqrt{N_{Data}}}$ from the
    fit.  \label{fitpnpipi} }
}
\end{figure}

The fits to the $\pnpim$ and $\pnpip$ invariant mass distributions are
shown in Fig.~\ref{fitpnpi} (a) and~\ref{fitpnpi} (b), respectively.
The corresponding fits to the $\pnpipim$ and $\pnpipip$ invariant mass
distributions are shown in Figs.~\ref{fitpnpipi} (a) and (b),
respectively.


MC samples for signal events have been generated to obtain the
relevant detection efficiencies.  In the underlying event generators,
for the decay $\psip\to\gamma\chicj$ an angular distribution
proportional to 1+$\lambda\cos^{2}(\theta)$ has been assumed, where
$\theta$ is the angle between the direction of the radiative photon
and the positron beam, and $\lambda$ = 1, -1/3, 1/13 for $J$ = 0, 1,
2, respectively, in accordance with expectations of electric dipole
(E1) transitions.  Since in the case of
$\psip\to\gamma\chicj,\chicj\to\pnpim$, structures have been observed
in the $p\pi^-$, $\bar{n}\pi^-$, and $p\bar{n}$ invariant mass spectra
(see Sect.~\ref{sect:struct}),
the decays of $\chi_{cJ}$ into $\pnpim$ are generated taking these
structures and the polar angle distribution of the proton/neutron into
account.  The MC samples of $\chicj\to\pnpipim$ used to determine the
detection efficiencies were generated with a flat angular
distribution, although a $\rho^{\pm}$ intermediate state is seen.  To
estimate the systematic uncertainty associated with the missing
$\rho^{\pm}$ intermediate state, a MC production of $\chicj\to
p\bar{n}\rho^-$ with the correct angular distribution of
$\rho^{\pm}\to\pi^{\pm}\pi^0$ has been generated (see
Sect.~\ref{sect:syst}).

The $\chiczero$, $\chicone$ and $\chictwo$ MC samples are finally
weighted with the amplitudes observed in data, and the same fitting
process as that for data is performed to the mixed MC sample. The
detection efficiencies of $\chicj$ are calculated by
$\varepsilon_{cJ}=N^{fit}_{cJ}/N^{gen}_{cJ}$, where $N^{fit}_{cJ}$ is
the number of $\chicj$ events extracted from the fit, and
$N^{gen}_{cJ}$ is the number of generated $\chicj$ events.

\section{Branching fractions}
\label{sect:bf2}

\subsection{Branching fractions of $\chicj\to\pnpim$}

The branching
fractions of $\chicj\to\pnpim$ are calculated according to:
\begin {equation}
 \label{eq2}
  \mathcal{B}(\chicj\to\pnpim)=\frac {N_{sig}}{N_{\psip}\times \mathcal{B}(\psip\to\gamma\chicj)\times
\varepsilon_{cJ}},
\end {equation}
where $N_{sig}$ is the number of signal events extracted from the fit
to the invariant mass distribution, $N_{\psip}$ is the number of
$\psip$ events, $\mathcal{B}(\psip\to\gamma\chicj)$ is the branching
fraction of $\psip\to\gamma \chicj$ as quoted in the PDG~\cite{pdg},
and $\varepsilon_{cJ}$ is the detection efficiency. The results are
summarized in the left column in Table~\ref{tabfitpnpi}. The same
calculation for the charge conjugate channel is performed, and the
results are summarized in the right column.

\begin {table}[htp]
  \begin {center}{ \caption{ The number of signal events $N_{sig}$,
        the detection efficiency $\varepsilon_{cJ}$, and the branching
        fractions of $\chicj\to\pnpim $, where the errors are
        statistical only.}
 \label{tabfitpnpi}
}
\begin {tabular}{l |c c c |c c c} \hline \hline
\footnotesize
                      & \multicolumn{3}{c|}{$\chicj\to\pnpim$}& \multicolumn{3}{c}{$\chicj\to\pnpip$} \\
                      & $\chiczero$      & $\chicone$ & $\chictwo$ & $\chiczero$ & $\chicone$ & $\chictwo$ \\ \hline

  $N_{sig}$             &~~ 5150$\pm$102    &~~ 1412$\pm$58    &~~ 3309$\pm$79    &~~ 5808$\pm$121   &~~ 1625$\pm$73    &~~ 3732$\pm$89    \\
  $\varepsilon_{cJ}$ (\%)    &~~ 38.6$\pm$0.2    &~~ 35.9$\pm$0.3   &~~ 39.2$\pm$0.2   &~~ 40.9$\pm$0.2    &~~ 40.7$\pm$0.3  &~~ 41.2$\pm$0.2   \\
  $\mathcal{B}$  $(10^{-3})$ &~~ 1.30$\pm$0.03   &~~ 0.40$\pm$0.02  &~~ 0.91$\pm$0.02  &~~ 1.38$\pm$0.03   &~~ 0.41$\pm$0.02  &~~ 0.98$\pm$0.02   \\ \hline \hline
\end {tabular}
\end {center}
\end {table}

\subsection{Branching fractions of $\chicj\to\pnpipim$}

Considering the branching fraction of $\pi^0\to\gamma\gamma$, the
branching fraction of $\chicj\to\pnpipim$ is calculated according to:
\begin {equation}
\label{eq3}
  \mathcal{B}(\chicj\to\pnpipim)=\frac {N_{sig}}{N_{\psip}\times
  \mathcal{B}(\psip\to\gamma\chicj)\times \mathcal{B}(\pi^{0}\to\gamma\gamma)\times{\varepsilon_{cJ}}}.
\end {equation}
The results are summarized in the left column in
Table~\ref{tabfitpnpipi}. The corresponding results for the charge
conjugate channel are also shown in the right column.
\begin {table}[htp]
  \begin {center}{ \caption {The number of signal events $N_{sig}$,
        the detection efficiency $\varepsilon_{cJ}$ and the branching
        fractions of $\chicj\to\pnpipim$, where the errors are
        statistical only.}
   \label{tabfitpnpipi}
   }
  \begin {tabular}{l |c c c |c c c}  \hline \hline
  \footnotesize
                        & \multicolumn{3}{c|}{$\chicj\to \pnpipim$}& \multicolumn{3}{c}{$\chicj\to \pnpipip$} \\
                        & $\chiczero$      & $\chicone$ & $\chictwo$ & $\chiczero$ & $\chicone$ & $\chictwo$ \\ \hline
  $N_{sig}$             &~~ 2480$\pm$85  &~~ 1082$\pm$52 &~~ 2128$\pm$62  &~~ 2757$\pm$94    &~~ 1261$\pm$60   &~~ 2352$\pm$69 \\
  $\varepsilon_{cJ}$ (\%)    &~~ 10.4$\pm$0.1 &~~ 10.4$\pm$0.2 &~~ 9.8$\pm$0.1 &~~ 12.2$\pm$0.1   &~~  12.3$\pm$0.2 &~~11.2$\pm$0.1 \\
  $\mathcal{B}$ $(10^{-3})$&~~ 2.36$\pm$0.08 &~~ 1.08$\pm$0.05 &~~ 2.38$\pm$0.07  &~~ 2.23$\pm$0.08 &~~ 1.06$\pm$0.05 &~~ 2.30$\pm$0.07 \\ \hline \hline
\end {tabular}
\end {center}
\end {table}

\section{ ESTIMATION OF SYSTEMATIC UNCERTAINTIES}
\label{sect:syst}
Several sources of systematic uncertainties are considered in the
measurement of the branching fractions.  These include differences
between data and the MC simulation for the tracking algorithm, the
particle identification (PID), photon detection, the kinematic fit,
the requirement on the angle $\alpha$, $\pi^0$ reconstruction, the
fitting procedure, and the number of $\psip$ events. Also possible
imperfections in the description of intermediate resonances in the MC
are considered.

\emph{a. Tracking and PID~~}
  The uncertainties from tracking efficiency and PID are investigated
  using an almost background-free control sample of
  $J/\psi\to p\bar{p}\pi^{+}\pi^{-}$ from $(225.3\pm 2.8)\times 10^8$ $J/\psi$ decays~\cite{jpsinumber}.
  The tracking efficiency is calculated with
  $\epsilon=N_{full}/N_{all}$, where $N_{full}$ is the number of
  events with all final tracks reconstructed successfully and
  $N_{all}$ is the number of events with three of them reconstructed
  while one track is missing.  The PID efficiency is the ratio of the
  number of selected events with and without PID.  Both efficiencies
  are studied for pions and protons (antiprotons) as a function of
  transverse momentum and $\cos\theta$.  The data - MC simulation
  difference for the tracking efficiency is estimated to be 1\% per
  track. Therefore, a 2\% uncertainty is taken for two-track events.
  For the PID efficiency, a 2\% difference between data and MC is
  found for antiprotons, and 1\% for any other charged particle.
  Therefore, 2\% (3\%) is taken as the systematic uncertainty for the
  final states including $p\pi^-$ ($\bar{p}\pi^+$).

  \emph{b. Photon detection~~} The uncertainty due to photon detection
  and photon conversion is 1\% per photon. This value is determined
  from a study using clean control samples, such as
  $\jpsi\to\rho^{0}\pi^{0}$ and
  $e^{+}e^{-}\to\gamma\gamma$. Therefore, a 1\% uncertainty is taken
  for the $\psip\to\gamma\chicj\to\gamma\pnpim$ channel, while for the
  $\psip\to\gamma\chicj\to\gamma\pnpipim$ channel, with 3 photons in
  the final state, a 3\% uncertainty is taken.

  \emph{c. Kinematic fit~~} The systematic uncertainty stemming from
  the 1C kinematic fit is investigated using $\jpsi\to\pnpim$ and
  $\jpsi\to\pnpipim$ events, where $\bar{n}$ is treated as a missing
  particle with a mass of 0.938 GeV/$c^2$.  To obtain the systematic
  uncertainty associated with the fit, pure control samples of
  $\jpsi\to\pnpim$ and $\jpsi\to\pnpipim$ are selected, and the 1C
  kinematic fit is applied to both charge states, both for data and
  MC. The efficiency of the 1C kinematic fit is estimated calculating
  the ratio of the number of events with and without the kinematic
  fit.  From the two charge conjugate states, the larger difference
  between data and MC is taken as the systematic uncertainty.  We
  assign an uncertainty of 2.9\% for the kinematic fit in the case of
  $\pnpim(\pi^{0})$ and 2.7\% for $\pnpip(\pi^{0})$.

  \emph{d. $Angle$ $\alpha$ ~~} Another source of systematic
  uncertainty is the requirement on the angle $\alpha_{\gamma \bar{n}}
  <15^{\circ}$. This uncertainty is studied using the $\jpsi\to\pnpim$
  sample and an uncertainty of 1.8\% is assigned for this item.

  \emph{e. $\pi^{0}$ reconstruction~~} The uncertainty from the
  $\pi^0$ reconstruction is determined with a high purity and high
  statistics sample of $J/\psi\to\pi^{+}\pi^{-}\pi^{0}$.  The
  $\pi^{0}$ selection efficiency is determined by counting the number
  of $\pi^0$ candidates in the $\pi^{+}\pi^{-}$ recoiling mass
  distribution with and without the standard $\pi^{0}$ selection (1C
  kinematic fit). The data - MC simulation difference has been
  measured to be 0.7\%, and it has been verified that there is no
  dependence of this value from the $\pi^0$ momentum and $\pi^0$ polar
  angle.

  \emph{f. Uncertainty from intermediate states~~} As mentioned above,
  in $\chicj\to p\bar{n}\pi^{-}$, obvious $N^\ast$ intermediate states
  and a $p\bar{n}$ threshold enhancement are observed in the invariant
  mass spectra of $p\pi^{-}$ ($\bar{n}\pi^{-}$) and $p\bar{n}$
  (Fig.~\ref{Nstar-plot}). To account for these structures in the
  determination of the detection efficiency for $\chicj\to
  p\bar{n}\pi^{-}$, MC samples were produced including these
  structures at the generator level.  To determine the systematic
  error associated with this procedure, an alternative method is used,
  where the efficiency, including the effect of these structures for
  $\chicj\to p\bar{n}\pi^{-}$, is determined by re-weighting MC events
  generated according to phase space by the ratio of data and MC
  events in the two-dimensional distribution of $p\pi^{-}$ versus
  $\bar{n}\pi^{-}$ invariant mass.  The difference in the detection
  efficiencies determined by the two methods is assigned as the
  systematic uncertainty.  In $\chicj\to\pnpipim$, a strong
  $\rho^{\pm}$ signal is observed in the $\pi^{\pm}\pi^0$ invariant
  mass spectrum (Fig.~\ref{rho}), while MC samples used to estimate
  the detection efficiencies are generated assuming phase space only.
  To study the effect of the intermediate state on the efficiencies, a
  sample of $\chicj\to p\bar{n}\rho^-$ events is generated and
  analyzed.  The efficiency difference between the two MC samples with
  and without a $\rho^\pm$ intermediate state,
  ($\varepsilon_{pn\pi\pi^0}-\varepsilon_{pn\rho}$)/$\varepsilon_{pn\pi\pi^0}$,
  is taken as the systematic uncertainty.

  \emph{g. Fitting procedure~~} As described above, the yields of the
  $\chi_{cJ}$ signal events are derived from fits to the invariant
  mass spectra of $\pnpim$ and $\pnpipim$.  To evaluate the systematic
  uncertainty associated with the fitting procedure, we have studied
  the following aspects: (i) {\it Fitting range:} In the nominal fit,
  the mass spectra of $\pnpim$ and $\pnpipim$ are fitted from 3.30
  GeV/$c^2$ to 3.60 GeV/$c^2$ and from 3.30 GeV/$c^2$ to 3.64
  GeV/$c^2$, respectively. We have changed these intervals to
  3.32$-$3.60 GeV/$c^2$ in the case of $\pnpim$ and 3.32$-$3.62
  GeV/$c^2$ for $\pnpipim$ events.  The differences in the finally
  obtained branching fractions by changing the fit intervals, are
  taken as the systematic uncertainties associated with the fit
  intervals.  (ii) {\it Signal lineshape:} The partial width for an
  E1/M1 radiative transition is proportional to the cube of the
  radiative photon energy ($E^{3}_{\gamma}$), which leads to a
  diverging tail in lower mass region.  Two damping functions have
  been proposed by the KEDR~\cite{KEDR} and the CLEO ~\cite{CLEO}
  collaborations and have been used in addition to the standard
  approach in the formula describing the fit to the signal
  lineshape. Differences with respect to the fit not taking into
  account this damping factor have been observed, and the greater of
  the two differences is taken as the systematic uncertainty
  associated with the signal lineshape.  (iii) {\it Mass resolution
    parameterization:} A single Gaussian formula instead of the
  modified Gaussian formula (Formula 2) has been used to describe the
  instrumental mass resolution. The resulting differences for the
  final branching fractions are taken as the systematic uncertainties.
  (iv) {\it Mass resolution:} Studies have shown that the $\chicj$
  mass resolutions, as simulated by MC, are underestimated.  To
  evaluate the systematic effects associated with this aspect, the
  invariant masses of $\pnpim$ and $\pnpipim$ in the MC samples are
  smeared with a Gaussian function, where the width of this Gaussian
  depends on the invariant mass as well as on the channel.  The same
  fitting processes as in the nominal cases are performed on the
  smeared mass spectra of $\pnpim$ and $\pnpipim$, and the detection
  efficiencies are recalculated. The efficiency difference between the
  smeared and unsmeared case is taken as the systematic uncertainty.
  (v) {\it Background shape:} To estimate the uncertainties due to the
  background parameterizations, a second order instead of a third
  order Chebychev polynomial is applied in the fitting. Again, the
  difference between the two cases is used as an estimate of the
  systematic uncertainty.

  \emph{h. Other systematic uncertainties~~} The number of $\psip$
  events is determined from an inclusive analysis of $\psip$ hadronic
  events and an uncertainty of 4\%~\cite{totaln} is associated to
  it. The uncertainties due to the branching fractions of
  $\psip\to\gamma\chicj$ and $\pi^{0}\to\gamma\gamma$ are taken from
  the PDG~\cite{pdg}.  A small uncertainty due to the statistical
  error of the efficiencies is also considered.

  In Table~\ref{systematic} a summary of all contributions to the
  systematic error is shown.  In each case, the total systematic
  uncertainty is obtained by adding the individual contributions in
  quadrature.

\begin {table}[htp]
\begin {center}
  \caption {Summary of systematic errors (in \%) for the branching fraction measurements
    of $\chicj\to\pnpim$ and $\chicj\to\pnpipim$. }
  \label{systematic}
\begin {tabular}{l | c c c | c c c | c c c | c c c} \hline \hline
 \footnotesize  & \multicolumn{3}{c|}{$\chicj\to\pnpim$}   & \multicolumn{3}{c|}{$\chicj\to\pnpip$}
                & \multicolumn{3}{c|}{$\chicj\to\pnpipim$} & \multicolumn{3}{c}{$\chicj\to\pnpipip$}\\
 \footnotesize  &~~$\chiczero$~~&~~$\chicone$~~&~~$\chictwo$~~&~~$\chiczero$~~&~~$\chicone$&$\chictwo$~~
                &~~$\chiczero$~~&~~$\chicone$~~&~~$\chictwo$~~&~~$\chiczero$~~&~~$\chicone$&$\chictwo$~~\\
 \hline
MDC tracking                             &2.0&2.0&2.0&2.0&2.0&2.0&2.0&2.0&2.0&2.0&2.0&2.0\\
PID                                  &2.0&2.0&2.0&3.0&3.0&3.0&2.0&2.0&2.0&3.0&3.0&3.0\\
Photon detection                     &1.0&1.0&1.0&1.0&1.0&1.0&3.0&3.0&3.0&3.0&3.0&3.0\\
Kinematic fit                        &2.9&2.9&2.9&2.7&2.7&2.7&2.9&2.9&2.9&2.7&2.7&2.7\\
$\alpha_{\gamma\bar{n}}<15^{\circ}$        &1.8&1.8&1.8&-  &-  &-  &1.8&1.8&1.8&-  &-  &- \\
$\pi^{0}$ reconstruction              & - & - & - &-  &-  &-  &0.7&0.7&0.7&0.7&0.7&0.7\\
Intermediate states   &5.3 &6.2 & 8.2 &4.0 &1.2 &5.0 &3.0&1.7&0.7&1.5&1.8&2.3\\
Fit range                            &0.3&0.1&0.1&0.6&0.2&0.1&0.0&1.6&1.1&0.4&2.4&0.8\\
Signal lineshape                     &1.4&2.4&1.1&0.9&5.4&1.4&0.8&2.5&1.1&1.5&1.7&0.8\\
Resolution para.                     &1.9&4.6&3.0&3.1&5.3&3.1&2.1&5.5&1.7&0.4&4.9&2.2\\
Resolution diff.                    &0.3&1.4&1.3&0.5&0.2&0.7&1.9&1.9&1.0&0.8&2.4&1.8\\
Background shape                     &1.7&5.8&1.2&0.3&1.1&0.2&1.3&1.6&1.1&1.1&2.4&0.9\\
N($\psip$)                             &4.0&4.0&4.0&4.0&4.0&4.0&4.0&4.0&4.0&4.0&4.0&4.0\\
$\mathcal{B}(\psip\to\gamma\chicj)$  &3.2&4.3&3.9&3.2&4.3&3.9&3.2&4.3&3.9&3.2&4.3&3.9\\
$\mathcal{B}(\pi^0\to\gamma\gamma)$  &-  &-  &-  &-  &-  &-  &0.03&0.03&0.03&0.03&0.03&0.03\\
MC statistics                         &0.2&0.3&0.2&0.2&0.3&0.2&0.1&0.2&0.1&0.1&0.2&0.1\\
\hline
Total                                  &9.1&12.5&11.5&8.6&10.8&9.5&8.6&10.6&8.3&7.9&10.6&8.8 \\
\hline
\hline
\end {tabular}
\end {center}
\end {table}

\section{SUMMARY}
Based on a data sample of 1.06$\times10^{8}$ $\psip$ events collected
with the BESIII detector, the branching fractions of $\chicj\to\pnpim$
and $\chicj\to\pnpipim$ are measured for $J=0,1,2$. The results are
summarized in Table~\ref{tabresults}, where for each branching
fraction the first error is statistical and the second systematic. The
product branching fractions of $\mathcal{B}(\psip\to\gamma\chi_{cJ})
\times \mathcal{B}(\chi_{cJ}\to\pnpim)$ and
$\mathcal{B}(\psip\to\gamma\chi_{cJ})\times
\mathcal{B}(\chi_{cJ}\to\pnpipim)$ are also summarized in
Table~\ref{extra}.  For $\chiczero\to\pnpim$ and $\chictwo\to\pnpim$,
the results are consistent with the world average values within one
standard deviation, while the precision is improved significantly. For
the other $\chi_{cJ}$ decay modes, the branching fractions are
measured for the first time. A comparison of individual branching
fraction shows good agreement between charge conjugate channels.  The
measurements improve the existing knowledge of the $\chi_{cJ}$ states
and may provide further insight into their decay mechanisms. Based on
the results of this work, detailed studies concerning the intermediate
states may follow in the future.

\begin {table}[htp]
  \begin {center} {\caption { Summary of branching fractions for
        $\chicj\to\pnpim$ and $\chicj\to\pnpipim$.  The first errors
        are statistical, and the second ones are systematic.  }
 \label{tabresults}
}
\begin {tabular}{ l c c c c } \hline \hline
\footnotesize                                         &     $\chiczero$               &   $\chicone$       &       $\chictwo$       \\ \hline
{$\mathcal{B}$($\chicj\to\pnpim$)  ($10^{-3}$)~} &~ 1.30$\pm$0.03$\pm$0.12&~~ 0.40$\pm$0.02$\pm$0.05&~~ 0.91$\pm$0.02$\pm$0.10 \\
{$\mathcal{B}$($\chicj\to\pnpip$)  ($10^{-3}$)~} &~ 1.38$\pm$0.03$\pm$0.12 &~~ 0.41$\pm$0.02$\pm$0.04&~~0.98$\pm$0.02$\pm$0.09 \\
{$\mathcal{B}$($\chicj\to\pnpipim$)($10^{-3}$)~} &~ 2.36$\pm$0.08$\pm$0.20 &~~1.08$\pm$0.05$\pm$0.12&~~2.38$\pm$0.07$\pm$0.20 \\
{$\mathcal{B}$($\chicj\to\pnpipip$)($10^{-3}$)~} &~ 2.23$\pm$0.08$\pm$0.18 &~~ 1.06$\pm$0.05$\pm$0.11&~~2.30$\pm$0.07$\pm$0.20 \\ \hline
{$\mathcal{B}$($\chicj\to\pnpim$)  ($10^{-3}$)~} (PDG~\cite{pdg})&~ 1.14$\pm$0.31 &        -              &~~ 1.10$\pm$0.40         \\\hline\hline
\end {tabular}
\end {center}
\end {table}

$~~~~~~~~~~~~~~~~~~~~~~~~~~~~~~~~~~~~~~~~~~~~~~~~~~~~~$

\begin {table}[htp]
  \begin {center} {\caption { Summary of the product branching
        fractions of $\psip\to\gamma\chicj$, $\chicj\to\pnpim$ and
        $\psip\to\gamma\chicj$, $\chicj\to\pnpipim$.  The first errors
        are statistical, and the second ones are systematic.  }
 \label{extra}
}
\begin {tabular}{ l c c c c } \hline \hline
\footnotesize                                         &     $\chiczero$               &   $\chicone$       &       $\chictwo$       \\ \hline
{$\mathcal{B}(\psip\to\gamma\chicj)\times\mathcal{B}$($\chicj\to\pnpim$)  ($10^{-4}$)~} &~ 1.26$\pm$0.02$\pm$0.11&~~ 0.37$\pm$0.02$\pm$0.04&~~ 0.80$\pm$0.02$\pm$0.09 \\
{$\mathcal{B}(\psip\to\gamma\chicj)\times\mathcal{B}$($\chicj\to\pnpip$)  ($10^{-4}$)~} &~ 1.34$\pm$0.03$\pm$0.11 &~~ 0.38$\pm$0.02$\pm$0.04&~~0.85$\pm$0.02$\pm$0.07 \\
{$\mathcal{B}(\psip\to\gamma\chicj)\times\mathcal{B}$($\chicj\to\pnpipim$)($10^{-4}$)~} &~ 2.29$\pm$0.08$\pm$0.18 &~~1.00$\pm$0.05$\pm$0.10&~~2.07$\pm$0.06$\pm$0.15 \\
{$\mathcal{B}(\psip\to\gamma\chicj)\times\mathcal{B}$($\chicj\to\pnpipip$)($10^{-4}$)~} &~ 2.16$\pm$0.07$\pm$0.16 &~~ 0.98$\pm$0.05$\pm$0.10&~~2.01$\pm$0.06$\pm$0.16 \\ \hline  \hline
\end {tabular}
\end {center}
\end {table}

  $~~~~~~~~~~~~~~~~~~~~~~~~~~~~~~~~~~~~~~~~~~~~~~~~~~~~~$
\section{ACKNOWLEDGMENTS}
The BESIII collaboration thanks the staff of BEPCII and the computing
center for their hard efforts. This work is supported in part by the
Ministry of Science and Technology of China under Contract
No. 2009CB825200; National Natural Science Foundation of China (NSFC)
under Contracts Nos. 10625524, 10821063, 10825524, 10835001, 10875113,
10935007, 11125525, 10979038, 11005109, 11079030; Joint Funds of the
National Natural Science Foundation of China under Contracts
Nos. 11079008, 11179007; the Chinese Academy of Sciences (CAS)
Large-Scale Scientific Facility Program; CAS under Contracts
Nos. KJCX2-YW-N29, KJCX2-YW-N45; 100 Talents Program of CAS; Research
Fund for the Doctoral Program of Higher Education of China under
Contract No. 20093402120022; Istituto Nazionale di Fisica Nucleare,
Italy; Ministry of Development of Turkey under Contract
No. DPT2006K-120470; U. S. Department of Energy under Contracts
Nos. DE-FG02-04ER41291, DE-FG02-91ER40682, DE-FG02-94ER40823;
U.S. National Science Foundation, University of Groningen (RuG) and
the Helmholtzzentrum fuer Schwerionenforschung GmbH (GSI), Darmstadt;
WCU Program of National Research Foundation of Korea under Contract
No. R32-2008-000-10155-0.


\end{document}